\documentclass[iop]{emulateapj}
\usepackage{graphicx}
\usepackage{multirow}
\usepackage{array}
\usepackage{subfigure}
\usepackage{hyperref}
\usepackage{microtype}
\usepackage[usenames]{color}

\begin{document}

\newcommand{\beq}{\begin{equation}}
\newcommand{\eeq}{\end{equation}}
\newcommand{\beqa}{\begin{eqnarray}}
\newcommand{\eeqa}{\end{eqnarray}}
\newcommand{\avg}[1]{\langle{#1}\rangle}
\newcommand{\hiMsun}{h^{-1} M_\odot}
\newcommand{\hiGpc}{h^{-1}{\rm Gpc}}
\newcommand{\hiMpc}{h^{-1}{\rm Mpc}}
\newcommand{\rmax}{r_\mathrm{max}}
\newcommand{\vmax}{v_\mathrm{max}}
\newcommand{\vtoday}{v_\mathrm{0}}
\newcommand{\vacc}{v_\mathrm{ac}}
\newcommand{\vpeak}{v_\mathrm{pk}}
\newcommand{\vcut}{v_\mathrm{cut}}
\newcommand{\Macc}{M_\mathrm{ac}}
\newcommand{\Mtoday}{M_\mathrm{0}}
\newcommand{\Mpeak}{M_\mathrm{pk}}
\newcommand{\Mvir}{M_\mathrm{vir}}
\newcommand{\Rvir}{R_\mathrm{vir}}
\newcommand{\Deltavir}{\Delta_\mathrm{vir}}
\newcommand{\kms}{\rm km\ s^{-1}}
\newcommand{\zhalf}{z_{\rm 1/2}}

\newcommand{\heidi}[1]{\textcolor{blue}{#1}}

\journalinfo{The Astrophysical Journal, {\rm 767:23 (14pp), 2013 April 10}}
\submitted{Received 2012 October 23; accepted 2013 February 21; published
2013 March 21}
\shortauthors{Wu et al.}

\shorttitle{Subhalos from Rhapsody Cluster Simulations}

\author{Hao-Yi Wu,$^{1,2}$  Oliver Hahn,$^{1,3}$  Risa H. Wechsler,$^1$
Peter S. Behroozi,$^1$ Yao-Yuan Mao$^1$ 
}
\affil{ 
$^1$Kavli Institute for Particle Astrophysics and Cosmology;
 Physics Department, Stanford University, Stanford, CA, 94305\\
 SLAC National Accelerator Laboratory, Menlo Park, CA, 94025\\
$^2$Physics Department, University of Michigan, Ann
Arbor, MI 48109; {\tt hywu@umich.edu}\\
$^3$Institute for Astronomy, ETH Zurich, CH-8093 Z\"urich, Switzerland}

\title{Rhapsody. II. Subhalo Properties and the Impact of Tidal Stripping \\From a Statistical Sample of Cluster-Size Halos}

\begin{abstract}
We discuss the properties of subhalos in cluster-size halos, using a
high-resolution statistical sample: the {\sc Rhapsody} simulations
introduced in \cite{Wu12}.  We demonstrate that the criteria applied
to select subhalos have significant impact on the inferred properties
of the sample, including the scatter in the number of subhalos, the
correlation between the subhalo number and formation time, and the
shape of subhalos' spatial distribution and velocity structure.  We
find that the number of subhalos, when selected using the peak maximum
circular velocity in their histories (a property expected to be
closely related to the galaxy luminosity), is uncorrelated with the
formation time of the main halo.  This is in contrast to the
previously reported correlation from studies where subhalos are
selected by the current maximum circular velocity; we show that this
difference is a result of the tidal stripping of the subhalos.  We
also find that the dominance of the main halo and the subhalo mass
fraction are strongly correlated with halo concentration and formation
history.  These correlations are important to take into account when
interpreting results from cluster samples selected with different
criteria.  Our sample also includes a fossil cluster, which is
presented separately and placed in the context of the rest of the
sample.
\end{abstract}


\section{Introduction}

The abundance and spatial distribution of galaxy clusters in the
universe have played an essential role in determining cosmological
parameters.  These properties are sensitive to cosmic expansion and
the large-scale structure growth rate, making clusters complementary to
other cosmological probes (see, e.g., \citealt{Allen11};
\citealt{Weinberg12} for recent reviews, and references therein).
Among multi-wavelength cluster surveys, optical surveys provide the
largest statistical power in terms of the number of identified
clusters; this number will dramatically increase with the next
generation of wide area surveys, including PanSTARRS\footnote{The
Panoramic Survey Telescope and Rapid Response System; \tt
http://pan-starrs.ifa.hawaii.edu/}, DES\footnote{The Dark Energy
Survey; \nolinebreak[5]\tt http://www.darkenergysurvey.org/}, Euclid\footnote{\tt
http://sci.esa.int/euclid/}, and LSST\footnote{The Large Synoptic
Survey Telescope; \tt http://www.lsst.org/}.  However, the precision
cosmology that can potentially be achieved will be limited by the
systematic effects involved, including cluster identification and
centering \cite[e.g.,][]{Rykoff12}, the normalization and scatter of
the richness--mass relation \cite[e.g.,][]{Rozo09Richness,Rozo11},
orientation and projection effects
\citep[e.g.,][]{Cohn07,WhiteM10,Erickson11}, cross-comparison with
multi-wavelength data \citep{Rozo12}, as well as uncertainties in
theoretical calibrations of halo statistics \citep{WuHY09b}.

To quantify the systematic effects inherent in the measurements of
galaxy clusters, it is essential to generate a simulated sample of
clusters that is comparable to the relevant observations.  A common
procedure is to use dark matter-only $N$-body simulations to predict the
distribution of dark matter particles and halos, and then relate halos
and subhalos to the observed galaxy clusters and their member galaxies
\citep[e.g.,][]{Kravtsov04,Zheng05}.  However, resolving subhalos in
cluster-size halos comparable to the observable limits and associating
them with galaxies presents additional challenges.  As described by
the hierarchical structure formation paradigm, subhalos accrete onto
the main halo through numerous merger events and have been
substantially influenced by the deep gravitational potential of the
main halo \citep[e.g.,][]{Ghigna98,Moore98,Moore99}.  Therefore,
simulating these subhalos requires high mass and force resolution
\citep[e.g.,][]{Klypin99}, improved halo finding
\citep[e.g.,][]{Onions12}, as well as a careful modeling of the
associated satellite galaxies \citep[e.g.,][]{Reddick12}.

To characterize the galaxy populations in galaxy clusters obtained
from deep wide surveys, it is necessary to simulate clusters with high
resolution (to resolve the galaxy content to observable limits) and in
a large cosmological volume (to obtain a statistical sample), which is
computationally challenging.  As discussed in Paper I \citep{Wu12}, in
order to achieve large sample size and high resolution simultaneously,
we have repeatedly applied a ``zoom-in'' or multi-resolution
simulation technique to develop the {\sc Rhapsody} sample, which
currently includes 96 halos of mass $10^{14.8\pm 0.05}\hiMsun$,
selected from a cosmological volume of side length 1 $\hiGpc$ and
re-simulated with mass resolution $1.3\times 10^8\hiMsun$.  This
sample is currently unique in terms of its sample size and resolution
and occupies a new statistical regime of cluster simulations
(see Figure 1 in Paper I).  In this second paper, we focus on the
subhalo population of the {\sc Rhapsody} clusters and give particular
attention to the impact of formation history and tidal stripping on
subhalos.

The impact of formation history on the observable properties of
clusters (e.g., galaxy number and distribution) is important because
it can provide extra information or, if not correctly taken into
account, introduce bias in cluster mass calibration and in
cosmological constraints that depend on the properties of galaxies in
clusters.  For example, \cite{WuHY08} have shown that if the richness
of a cluster (the number of galaxies in a cluster under a certain
selection criterion, used as a cluster-mass indicator) is correlated
with its formation time, then richness-selected clusters will be
impacted by assembly bias (i.e., early forming halos have a higher
halo bias; see, e.g.,
\citealt{Gao05,Harker06,Wechsler06,Croton07,Wetzel07,Hahn07,Hahn09}),
which will in turn impact cluster mass self-calibration and cause
systematic errors in the inferred cosmological parameters.
Cosmological studies which use information about the halo occupation
of galaxies (the number of galaxies inside a halo for a given halo
mass) also depend on an understanding of whether this occupation
depends on properties other than halo mass.  Therefore, it is
imperative to characterize these correlations with higher precision
using a statistical sample relevant for current and future surveys.

In this paper, we discuss how formation history impacts the subhalo
abundance, subhalo mass fraction, and the dominance of the main
halo over its subhalos.  In particular, we focus on the influence of
the specific criteria used to select subhalos from simulations on the
inferred properties of the subhalo population thus obtained.  One of
our main findings is that the correlation between subhalo number and
formation time sensitively depends on this selection criterion.  If we
use a subhalo selection criterion that is insensitive to the stripping
of dark matter particles, the subhalo number and halo formation time
are not correlated.  This result implies that both the halo occupation
and the cluster richness for halos of a given mass are not likely to
correlate with formation time.

This paper is organized as follows.  In Section~\ref{sec:catalogs}, we
briefly summarize our simulations and halo catalogs, as well as the
various subhalo selection criteria that we consider.  In
Section~\ref{sec:subs}, we present the statistics of subhalos and the shape
of their spatial distribution and velocity ellipsoid.  In
Section~\ref{sec:cor}, we discuss how formation time impacts observational
signatures of subhalos, including the subhalo mass fraction and the
dominance of the main halo.  In Section~\ref{sec:nocor}, we focus on the
impact of halo formation time and tidal stripping on the number of
subhalos.  We conclude in Section~\ref{sec:summary}.

\section{Halo Catalogs}\label{sec:catalogs}

The {\sc Rhapsody} sample includes 96 cluster-size halos of mass
$\Mvir = 10^{14.8 \pm 0.05} \hiMsun$, re-simulated from a cosmological
volume of 1 $h^{-3}{\rm Gpc}^3$.  Each halo has been simulated at two
resolutions: $1.3\times10^{8}\hiMsun$ (equivalent to $8192^3$
particles in this volume), which we refer to as ``{\sc Rhapsody 8K}''
or simply ``{\sc Rhapsody}''; and $1.0\times10^9 \hiMsun$ (equivalent
to $4096^3$ particles in this volume), which we refer to as ``{\sc
Rhapsody 4K}.''  The simulation parameters are summarized in Table 1
of Paper I.

All simulations in this work are based on a $\Lambda$CDM cosmology
with density parameters $\Omega_m = 0.25$, $\Omega_\Lambda= 0.75$,
$\Omega_b = 0.04$, spectral index $n_s = 1$, normalization $\sigma_8 =
0.8$, and Hubble parameter $h=0.7$.

\subsection{The Simulations}\label{sec:sim}

The implementation of {\sc Rhapsody} can be summarized as follows:
\begin{enumerate}
\item {\em Selecting the re-simulation targets.} We start from one of the 1
  $\hiGpc$ volumes (named ``{\sc Carmen}'') from the {\sc LasDamas} suite of
  simulations\footnote{\tt http://lss.phy.vanderbilt.edu/lasdamas/}
  and select halos in a narrow mass bin $10^{14.8\pm 0.05}\ \hiMsun$.
  We start from the center of this mass bin (which includes a total of $\sim$200 halos)
  and exclude those halos whose masses shift outside this  mass range after re-simulation.
  These leave us with 96 halos in the end.
\item {\em Generating initial conditions.} We use the multi-scale initial condition
  generator {\sc Music} \citep{HahnAbel11} to generate ``zoom'' initial conditions
  for each cluster with the second-order Lagrangian perturbation theory.
\item {\em Performing gravitational evolution.} We compute the non-linear evolution
of each cluster down to $z=0$ using the public version of {\sc Gadget-2} \citep{Springel05}.
\item {\em Identifying halos and subhalos.} We use the adaptive phase-space
  halo finder {\sc Rockstar} \citep{Behroozi11rs} to assemble catalogs of halos and
  subhalos at 200 output times. {\sc Rockstar} achieves a particularly high completeness of the subhalo 
  sample.
\item {\em Constructing merger trees.} We use the gravitationally consistent
  merger tree code by \cite{Behroozi11tree} to construct merger trees from the 
  halo/subhalo catalogs.
\end{enumerate}
We kindly refer the reader to Section~2 of Paper~I for more details on the simulations, the
halo and subhalo identification and merger tree generation, 
as well as the mean values and variances of the various key properties of the
main cluster halos in {\sc Rhapsody} (given in Table 2 of Paper~I).

\subsection{Subhalo Selection Methods}\label{sec:selection}
Subhalos in cluster-size halos are expected host the
observed satellite galaxies in clusters.  For each main halo in {\sc
  Rhapsody}, we consider the subhalos within its virial radius,
$\Rvir$ (based on the spherical overdensity calculated with
\citealt{BryanNorman98} $\Deltavir=\Delta_{94c}$ at $z$=0).  We
characterize each subhalo by the maximum circular velocity of the dark
matter particles associated with it, $\vmax$, defined at the radius
$r=\rmax$ that maximizes $\sqrt{{G M(<r)}/{r}}$: 
\beq \vmax =
\sqrt{\frac{G M(<\rmax)}{\rmax}} \ .
\label{eq:vmax}
\eeq 
This quantity is often used as a proxy for subhalo mass, since the
mass of a subhalo itself is typically not a well defined quantity in
the simulations due to ambiguities about how to separate subhalos from
the background density of the main halo.  We focus on $\vmax$ at
two different epochs during the evolution history of a subhalo:
\begin{itemize}
\item $\vtoday$: the value of $\vmax$ measured at $z=0$, a quantity
  related to the current subhalo mass. 
\item $\vpeak$: the highest $\vmax$ value in a subhalo's history, a
  quantity related to the highest subhalo mass in its entire history.
\end{itemize}
Since subhalos experience strong tidal stripping after their accretion
onto the main halo, the two are not identical.  The parameter $\vpeak$ is
more closely related to the luminosity and stellar mass of satellite
galaxies than $\vtoday$, because the stellar component of a galaxy is
denser and less easily stripped than the more extended dark matter
component. Even though a halo could lose dark matter particles at its
outskirts, the galaxy in its core can remain intact for a longer time.
In fact, the satellite galaxy may even continue to grow \citep[e.g.,][]{Wetzel12}.
Therefore, a quantity that is unaffected by stripping is expected to
provide a better proxy for the stellar mass of galaxies.

In particular, subhalo abundance matching models based on properties
that are less impacted by stripping have been shown to better agree
with observations. For example, \cite{NagaiKravtsov05} and
\cite{Conroy06} have shown that using $\vacc$ (i.e., $\vmax$ at the
time of accretion of the subhalo) when selecting subhalos better
reproduces the statistics of observed galaxies.  \cite{Reddick12} have
further demonstrated that an abundance matching model based on
$\vpeak$ provides a better fit than either $\vtoday$ or $\vacc$ to the
galaxy two-point correlation function and the conditional stellar mass
function for galaxies in groups from the Sloan Digital Sky Survey (SDSS).

In addition to $\vmax$, we also investigate the mass of subhalos at 
two different epochs:
\begin{itemize}
\item $\Mtoday$: the current subhalo mass, defined by the particles
  bound to the subhalo according to the implementation of {\sc Rockstar}. 
\item $\Mpeak$: the highest mass in a subhalo's entire assembly history.
\end{itemize}
In general, the difference between $\Mtoday$ and $\Mpeak$ is analogous
to the difference between $\vtoday$ and $\vpeak$,  except that the circular 
velocity is less affected by stripping than mass is.

It is important to note here that finite resolution in $N$-body
simulations leads to the ``overmerging'' effect
\citep[e.g.,][]{Klypin99}: small subhalos tend to fall below the
resolution limit before they merge with the central object. However,
the stellar component associated with subhalos is expected to survive
longer than the simulated subhalos.  A detailed discussion of the
resolution dependence of this effect and the associated completeness
limits of the subhalo populations will be presented in a separate
paper (H.-Y.~Wu et al., in preparation).

\section{Subhalo statistics and distributions at $z=0$} \label{sec:subs}
%%%%%%%%%%%%%%%%%%%%%%%%%%%%%%%%%%%%
\begin{figure*}
\includegraphics[width=0.66\columnwidth]{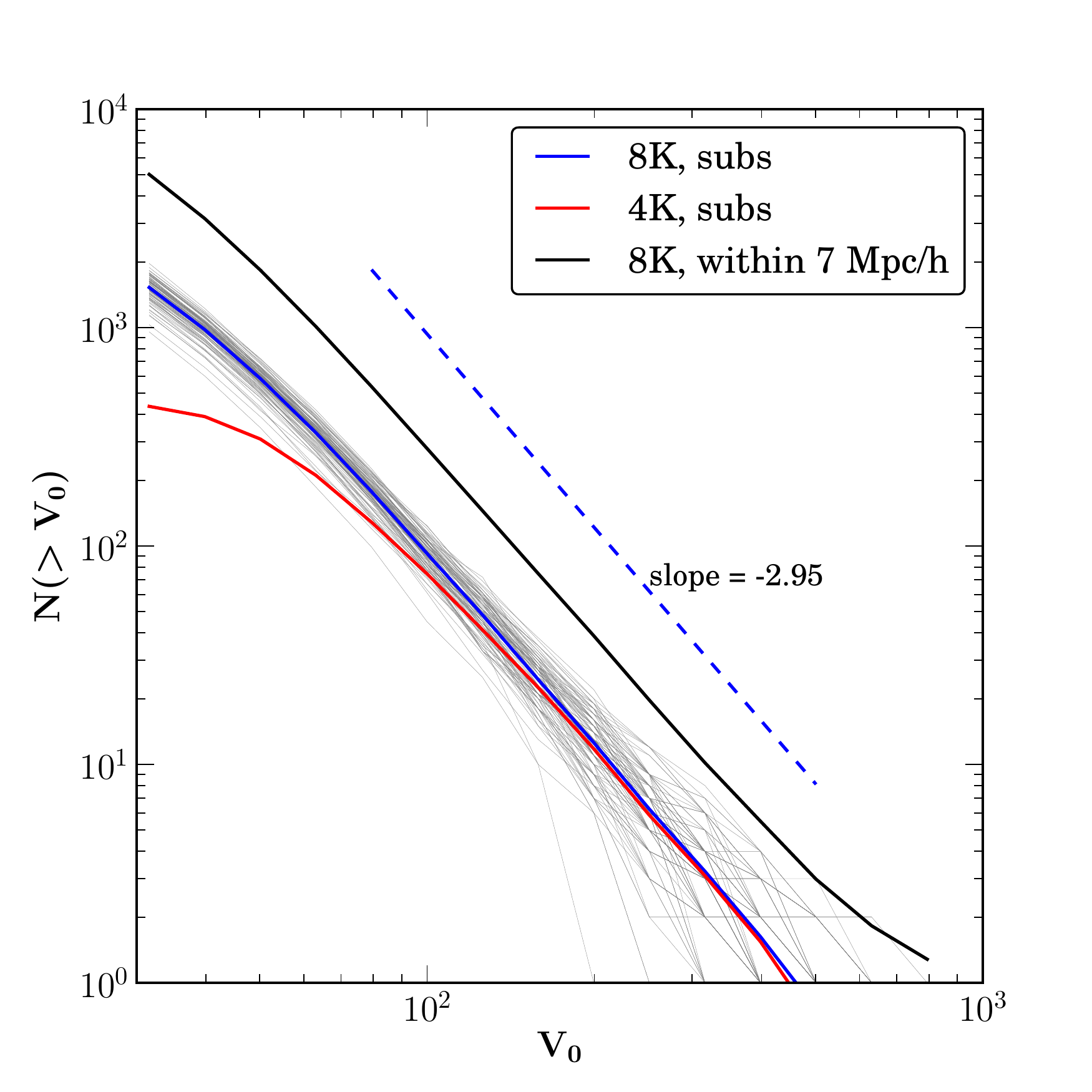}
\includegraphics[width=0.66\columnwidth]{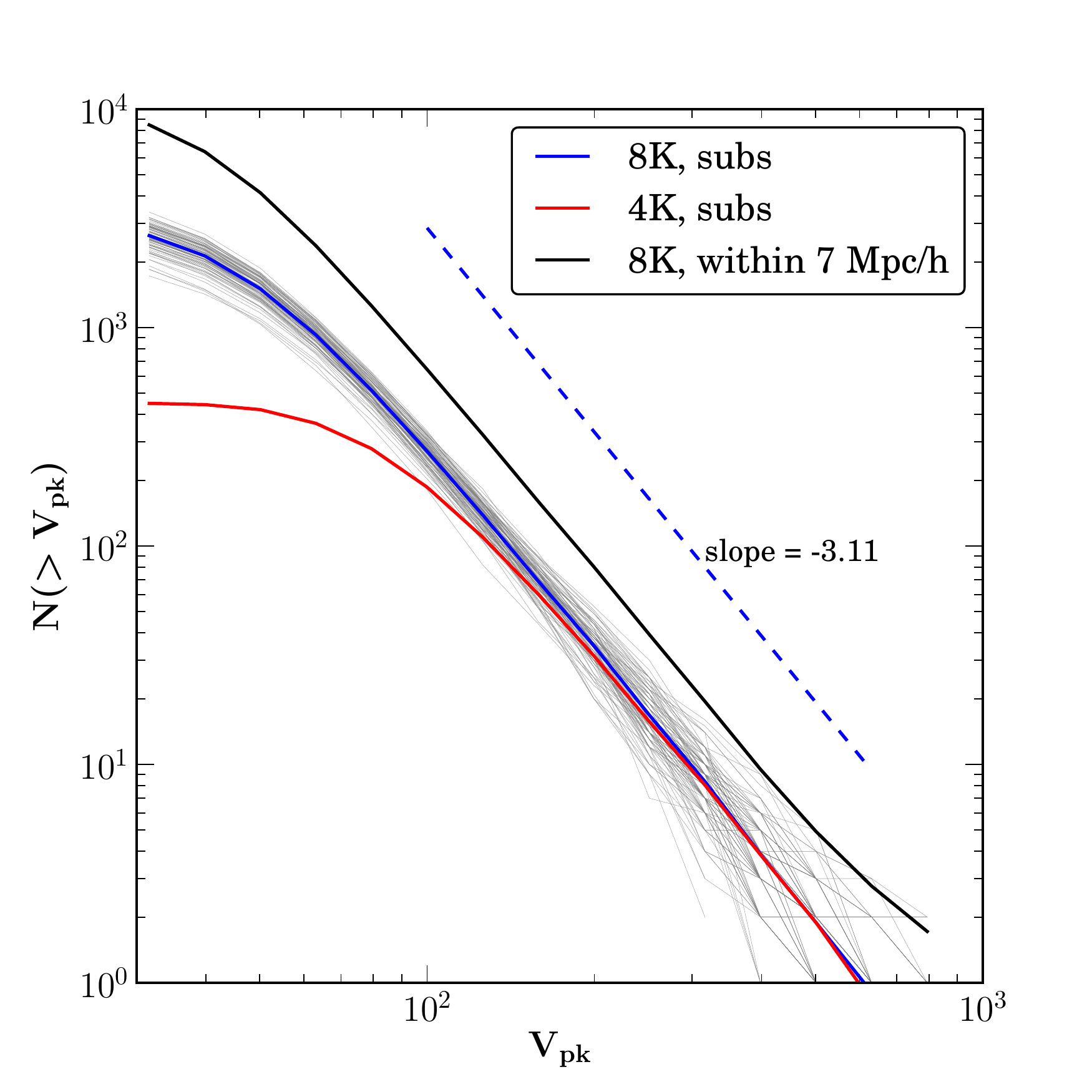}
\includegraphics[width=0.66\columnwidth]{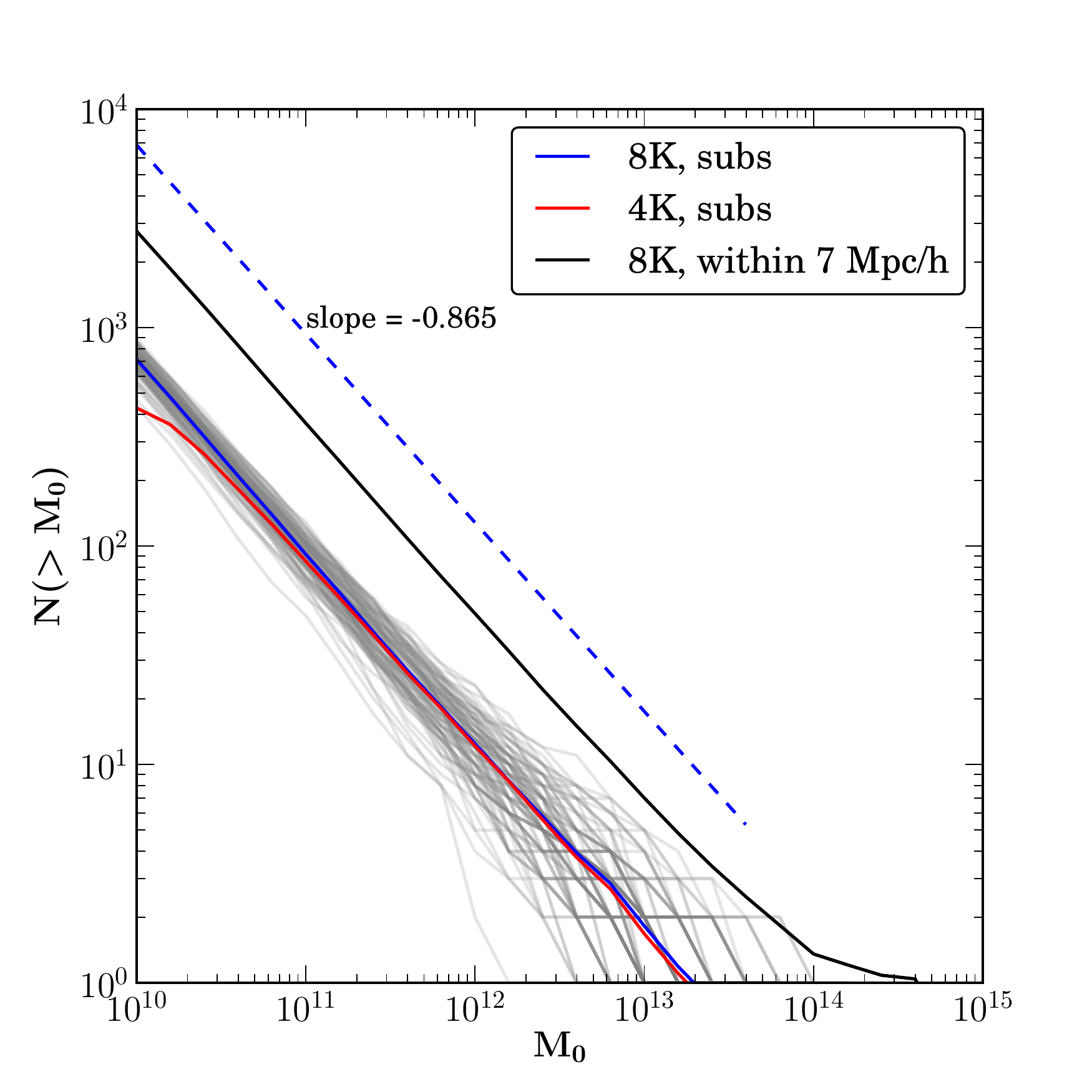}
\caption{Number of subhalos above a given threshold $\vtoday$
(left), $\vpeak$ (middle), and $\Mtoday$ (right).
The blue/red curves represent the mean number of subhalos within the
$\Rvir$ of {\sc Rhapsody} 8K/4K halos, while the black curves
represent all halos and subhalos within 7 $\hiMpc$ around the center
of the main halo in the re-simulation.  The thin gray curves in the
background show the subhalos for individual {\sc Rhapsody} 8K halos.
The blue dashed line indicates the slope of the 8K sample in the
regime where the subhalo sample is complete.  }
\label{fig:SHMF}
\end{figure*}
%%%%%%%%%%%%%%%%%%%%%%%%%%%%%%%%%%%%

In this section, we focus on the statistical properties of the
subhalos in {\sc Rhapsody}, as well as the shape of the subhalos' spatial
distribution and velocity ellipsoid.  In particular, we explore the
impact of resolution and of the various selection criteria described
in the previous section.

\subsection{Subhalo Mass Function}\label{sec:SHMF}

The mass function of subhalos has been shown to follow a power law for
low mass subhalos and an exponential cutoff for massive subhalos
\citep[e.g.,][]{Gao04, Angulo09, Giocoli10}.  In this section, we
investigate the validity of this form when using the various criteria
for subhalo selection .

Figure~\ref{fig:SHMF} shows the number of subhalos above a given
threshold of $\vtoday$ (left), $\vpeak$ (middle), and $\Mtoday$
(right).  The blue/red curves correspond to subhalos within $\Rvir$ of
{\sc Rhapsody} 8K/4K halos, while the transparent grey curves
correspond to individual halos in {\sc Rhapsody} 8K.  The black curves
correspond to all halos and subhalos within 7 $\hiMpc$ around the
center of the main halo, a region where re-simulated halos are well
resolved\footnote{To decide the size of the well-resolved ambient
region around the main halo, we compare halos in the re-simulated
region (composed of high-resolution particles only) with those in the
original {\sc Carmen} simulation.  At 7 $\hiMpc$, the re-simulated
regions recover the halo population in the corresponding region in the
{\sc Carmen} simulation.  In addition, at 7 $\hiMpc$, the number of
low-resolution particles is less than 4\%, although it varies from
halo to halo and is sometimes 0\%.}.  The blue dashed lines indicate
the best-fit power-law slopes of the distribution functions of the 8K
sample.  We find that the slope of our $\Mtoday$ function is slightly
shallower than \citet[][slope $-0.97$]{DeLucia04},
\citet[][slope $-0.935$]{Boylan-Kolchin10}, and \citet[][slope
$-0.94$]{Gao12}, which is plausibly attributed to the different mass
definition.  At the same time, our $\vtoday$ function is in good
agreement with the results of \citet[][slope $-2.98$]{Boylan-Kolchin10} 
and \citet[][slope $-3.11$]{Wang12} based on Milky Way-size halos.  The
large halo-to-halo scatter shown here has implications for comparisons
of satellite statistics in the Milky Way with simulations, given that
thus far these comparisons have been done with a small number of
simulated halos. Scatter in the properties of satellites between halos
likely reduces the current tension with observations of massive dwarf
galaxies \citep[e.g.,][]{PurcellZentner12}.  A statistical sample for
galactic subhalos from simulations, as well as a larger sample of
observed systems, is required to verify these results.

%%%%%%%%%%%%%%%%%%%%%%%%%%%%%%%%%%%%
\begin{figure*}%[t]
\centering
\includegraphics[width=\columnwidth]{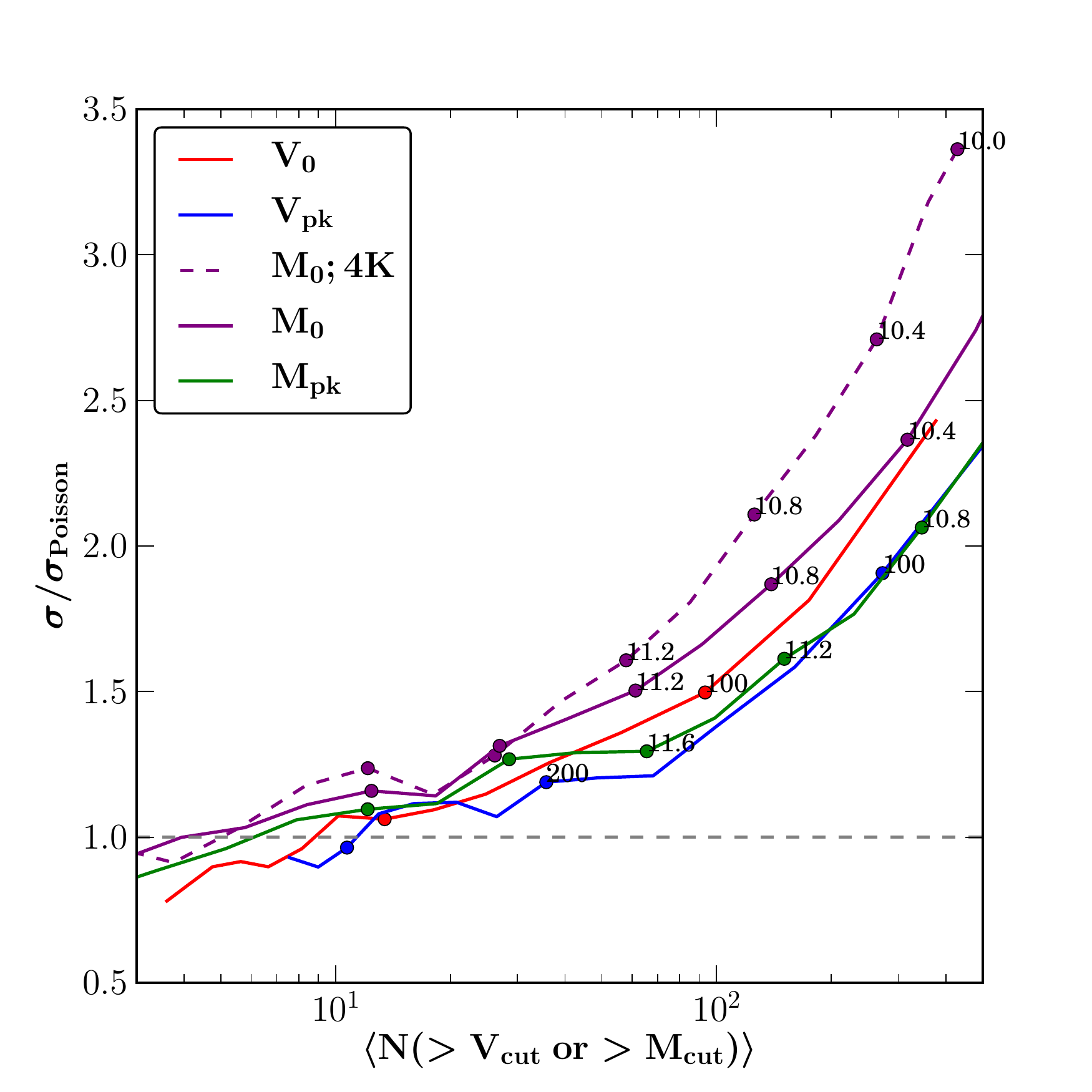}
\includegraphics[width=\columnwidth]{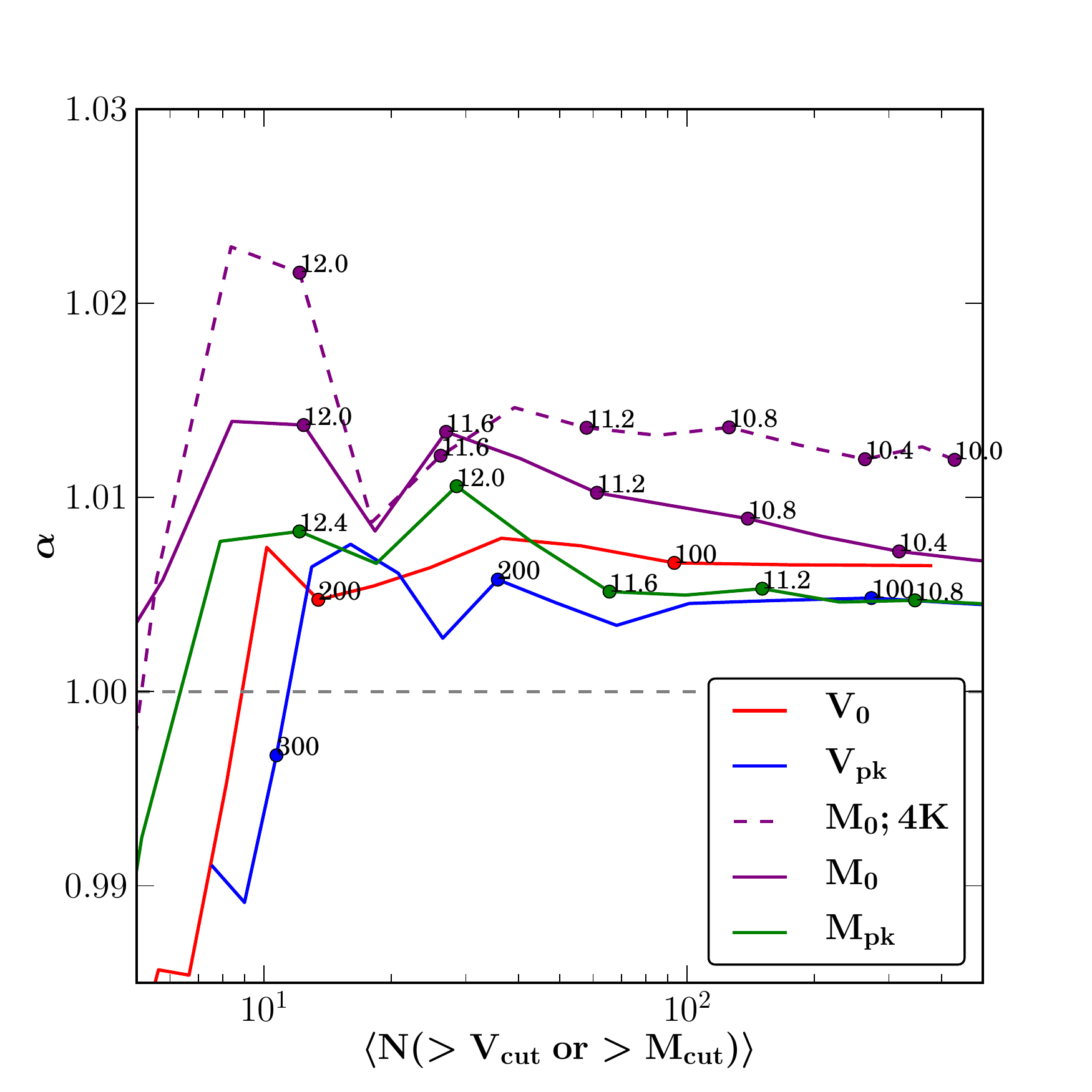}
\caption{Scatter of the number of subhalos for various selection
methods.  The left panel corresponds to the ratio between the sample
scatter and the Poisson scatter, while the right panel corresponds to
the second moment of the distribution.  The $x$-axis is the mean
number of subhalos for each selection method, and several of the
thresholds are marked on each curve.  Here we compare various cases:
(1) $\vtoday$ vs.~$\vpeak$ (red vs.~blue), (2) $\Mtoday$ vs.~$\Mpeak$
(purple vs.~green), (3) 4k vs.~8K (purple dashed vs.~purple solid).
In each pair, the former selection method leads to extra non-Poisson
scatter.  This trend indicates that both stripping of subhalos and
insufficient resolution can induce extra non-Poisson scatter and might
lead to a scatter greater than the values in observations.  For
$\vpeak$ selection, $\alpha$ = 1.005 for sufficiently large $\avg{N}$.
}
\label{fig:vcut_scatter}
\end{figure*}
%%%%%%%%%%%%%%%%%%%%%%%%%%%%%%%%%%%%
%%%%%%%%%%%%%%%%%%%%%%%%%%%%%%%%%%%%
\begin{figure*}%[t]
%\hspace*{-0.5in}
\includegraphics[width=3.5in]{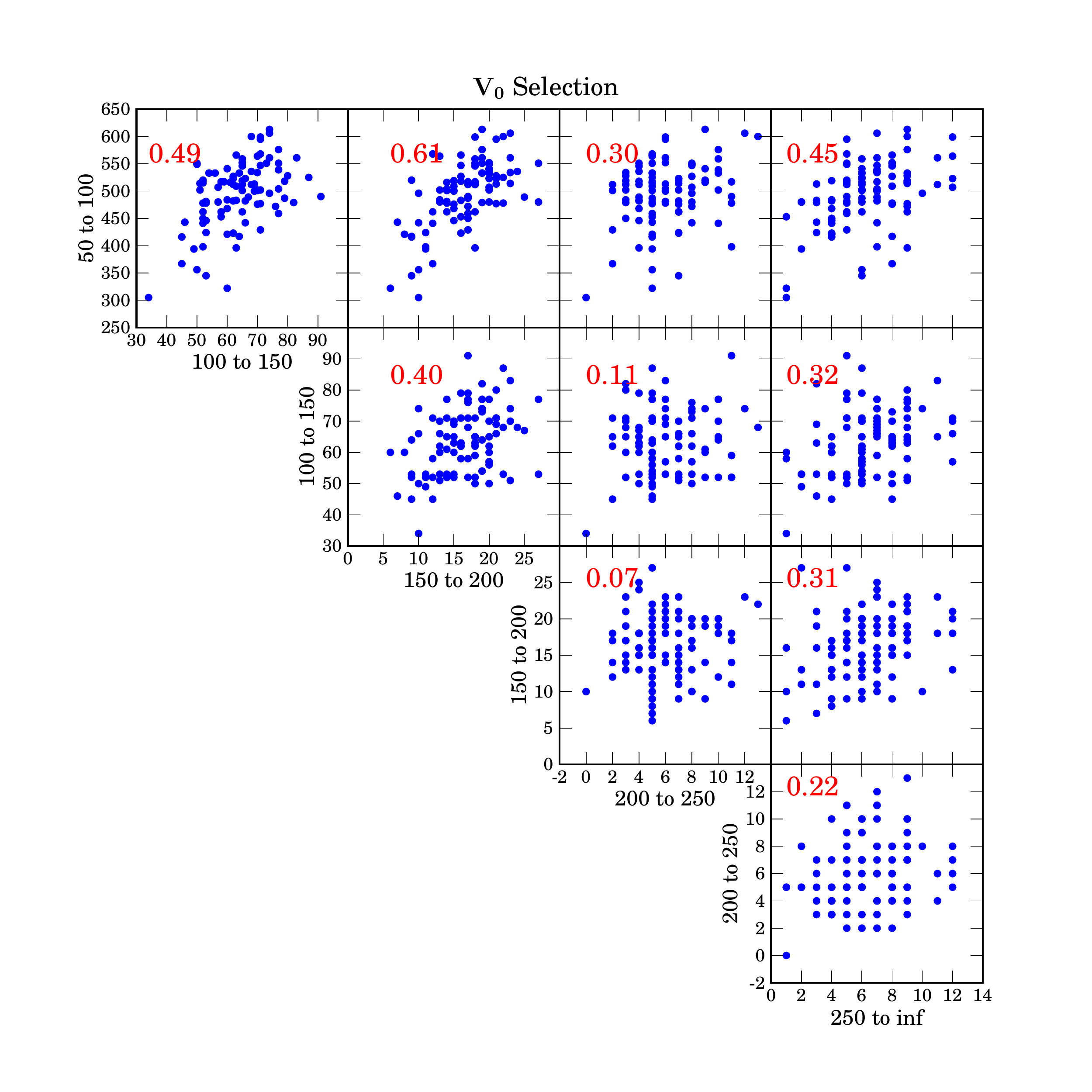}
\includegraphics[width=3.5in]{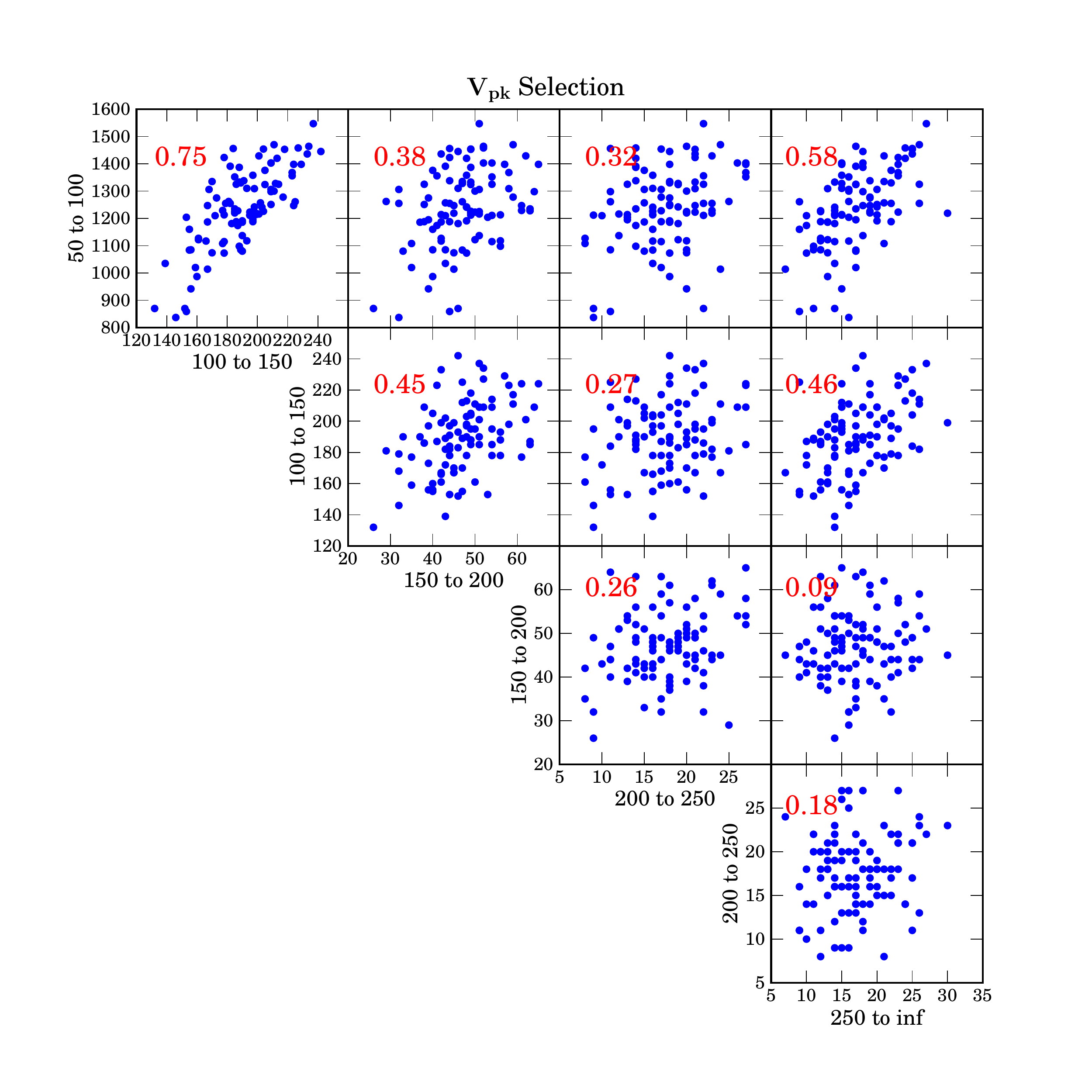}
\caption{Correlation between the number of subhalos in different bins.
Subhalos are binned by $\vtoday$ (left) or $\vpeak$ (right), with bin size 50 $\kms$, starting from 50 $\kms$.  The
subhalo numbers between different bins have moderate or weak
correlation.  }
\label{fig:50bins}
\end{figure*}
%%%%%%%%%%%%%%%%%%%%%%%%%%%%%%%%%%%%

The 4K and 8K subhalo mass functions deviate from a power law at
different values of $\vtoday$ and $\vpeak$, clearly indicating the
dependence of the completeness limit for all mass proxies on
resolution.  The gray curves demonstrate the significant scatter in
subhalo abundance from halo to halo in our sample.  In the next
section, we will explore how the scatter in the subhalo number depends
on mass, selection criterion, and resolution.

\subsection{Scatter of Subhalo Number}\label{sec:Poisson}

\cite{Boylan-Kolchin10} and \cite{Busha10} have shown that for
galactic halos ($M \le 10^{13.5} \hiMsun$), the distribution function
of the number of subhalos ($N$) deviates from the Poisson distribution
when $\avg{N}$ is large. We repeat this analysis for our sample of
halos of significantly higher mass.  Figure~\ref{fig:vcut_scatter}
shows the scatter of $N$ under different subhalo selection methods and
thresholds.  The left panel shows the ratio between the measured
scatter in the sample $\sigma = \sqrt{\rm{Var[N]}}$ and the Poisson
scatter $\sigma_{\rm Poisson} = \sqrt{ \avg{N}}$.  The right panel
presents the second moment of the subhalo number distribution
\beq
\alpha =\frac{\sqrt{\avg{N(N-1)}}}{\avg{N}} \ \mbox{($\alpha =1$ for
  Poisson)} \ .
\eeq
Both quantities are measures of how the distribution deviates from the
Poisson distribution.

In both panels, the $x$-axis corresponds to $\avg{N}$ for a given
selection threshold, allowing direct comparison between different
subhalo selections.  Each curve corresponds to a different selection
method, and several corresponding thresholds are marked on each curve
(in units of $\kms$ for maximum circular velocity and
$\log_{10}\hiMsun$ for mass).  For low thresholds or high $\avg{N}$,
the scatter deviates significantly from the Poisson scatter.  In
addition, $\alpha$ has a trend similar to that of $\sigma/\sigma_{\rm
Poisson}$ and only slightly deviates from unity.  As discussed in
\cite{Boylan-Kolchin10}, slight deviations of $\alpha$ from unity can
correspond to large deviations from the Poisson distribution.  We also
note that the additive boost of the variance has a similar trend as
$\alpha$, since $(\sigma^2 - \avg{N} )/\avg{N}^2 = \alpha^2-1$.

From Figure~\ref{fig:vcut_scatter}, our results for different
selection criteria can be summarized as follows:
\begin{itemize}
\item {\em $\vtoday$ versus $\vpeak$ (red versus blue).} $\vpeak$ selection gives
  less scatter and is closer to the Poisson distribution, indicating that stripping of
  subhalos can introduce extra non-Poisson scatter.
\item {\em $\Mtoday$ versus $\Mpeak$ (purple versus green).} $\Mpeak$ selection is
  closer to the Poisson distribution, which can also be understood
  with stripping.
\item {\em $\vpeak$ versus $\Mpeak$ (blue versus green):} similar.  Both
  properties are computed before a subhalo's infall and behave
  similarly.
\item {\em $\Mtoday$ $4K$ versus $8K$ (purple dashed versus purple solid).} 8K is closer to the
Poisson distribution, indicating that insufficient resolution can
introduce extra non-Poisson scatter.
\end{itemize}

In all cases presented here, the $\vpeak$ selection provides the
smallest non-Poisson scatter, with an asymptotic value of $\alpha$ =
1.005 for sufficiently large $\avg{N}$.  Our results suggest that
stripping and merging of subhalos can lead to extra non-Poisson
scatter.  The effects of stripping and merging are stronger for
subhalos selected with $\vtoday$ or $\Mtoday$ or with lower
thresholds, and are also stronger in low-resolution simulations.  In
these cases, the subhalo populations do not properly include
highly stripped or merged subhalos and tend to be more incomplete.
Therefore, they show larger non-Poisson scatter.  This trend can also
explain the difference between our results and the results in
\cite{Boylan-Kolchin10} ($\alpha=$1.02), who have slightly higher
resolution but selected subhalos by mass.

Given that the scatter depends on the selection threshold, we next
explore how well the number of subhalos in different bins are
correlated, i.e., how sensitive a richness estimator would be to a
different selection threshold. In Figure~\ref{fig:50bins}, we assign
subhalos into bins of 50 $\kms$ using $\vtoday$ (left) or $\vpeak$
(right), starting from 50 $\kms$ (we note that the first bin in either
case is incomplete, and the last bin includes all subhalos beyond 250
$\kms$).  We compare each pair of bins and find that the subhalo
counts are only weakly or moderately correlated between
bins\footnote{Throughout this work, we use rank correlation, which
makes our results insensitive to outliers.}.  This indicates that a
halo that is rich in massive subhalos is not necessarily also rich in
low-mass subhalos.  If the satellite galaxy populations in clusters in
different luminosity bins follow the statistical distribution shown
here and have such low covariances, they could potentially provide
independent information for mass calibration.  In addition, there is a
trend that the correlation between massive subhalos and less massive
subhalos increases with the decreased subhalo mass (the correlation
values tend to increase when we move upward in each column).  One
possible explanation is that massive subhalos tend to be accompanied
by a group of much smaller subhalos when they accrete onto the main
halo.

\subsection{Subhalo Spatial Distribution and Kinematics}\label{sec:sub_shape}
%%%%%%%%%%%%%%%%%%%%%%%
\begin{figure*}
\centering
\subfigure[]{\includegraphics[width=\columnwidth]{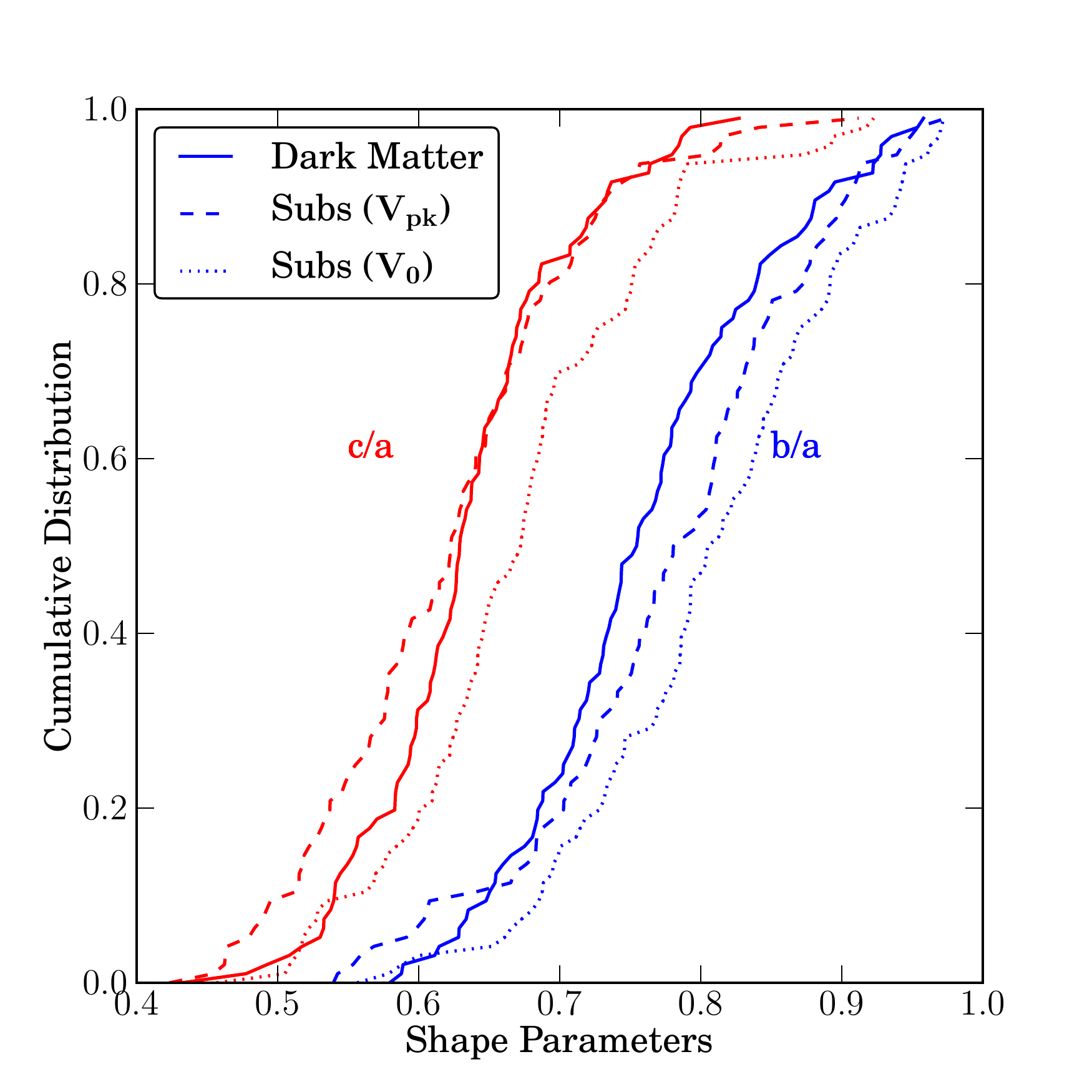}}
\subfigure[]{\includegraphics[width=\columnwidth]{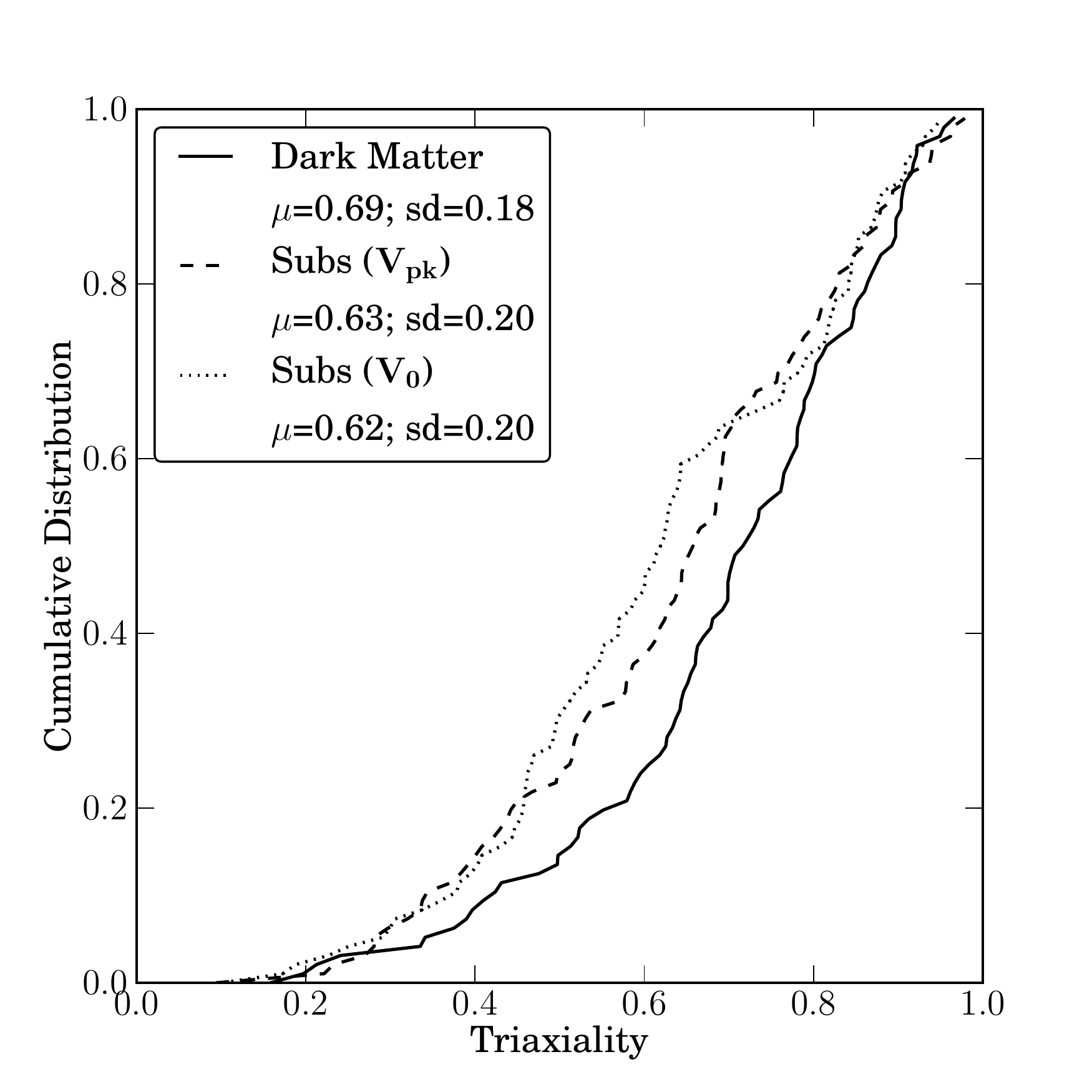}}\\
\subfigure[]{\includegraphics[width=\columnwidth]{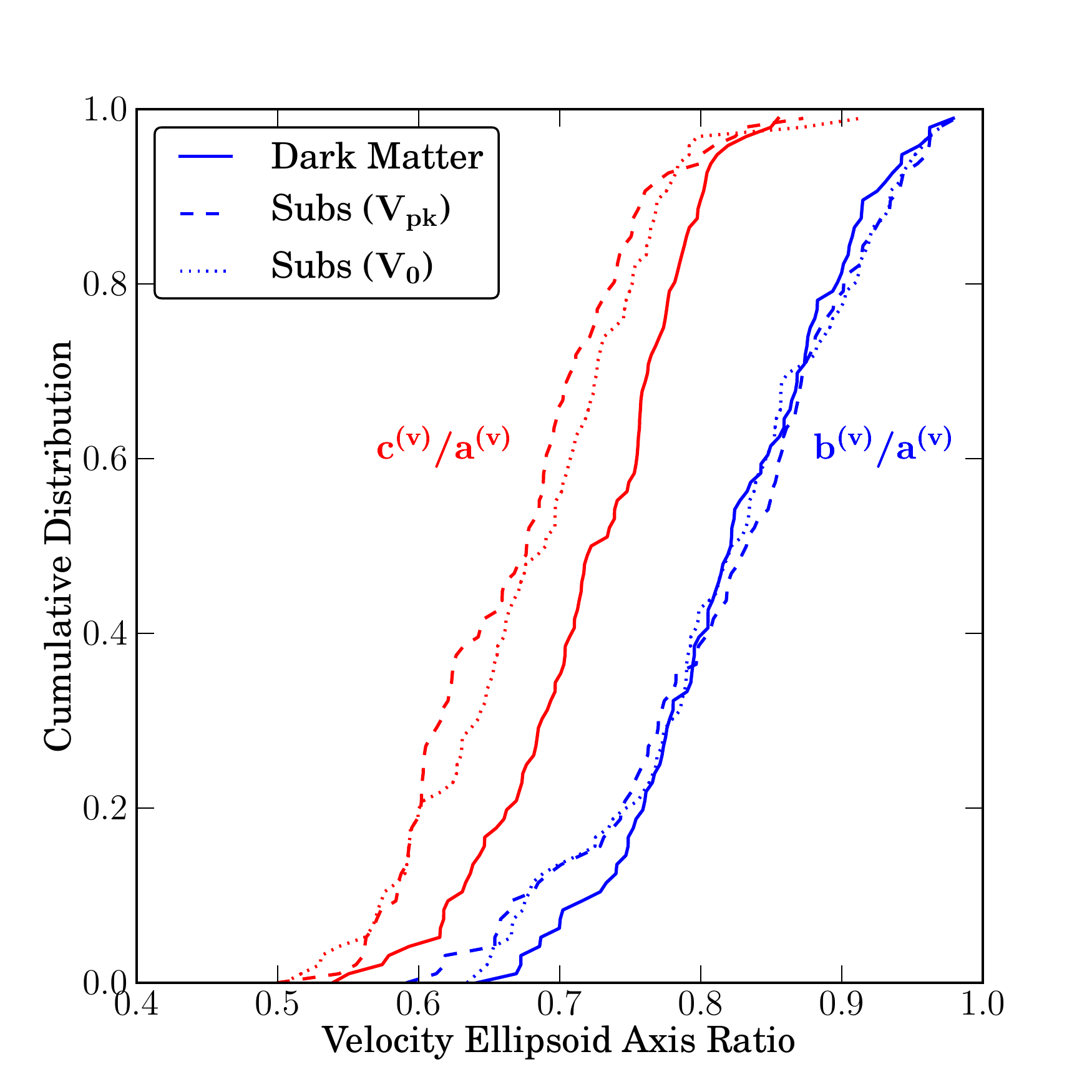}}
\subfigure[]{\includegraphics[width=\columnwidth]{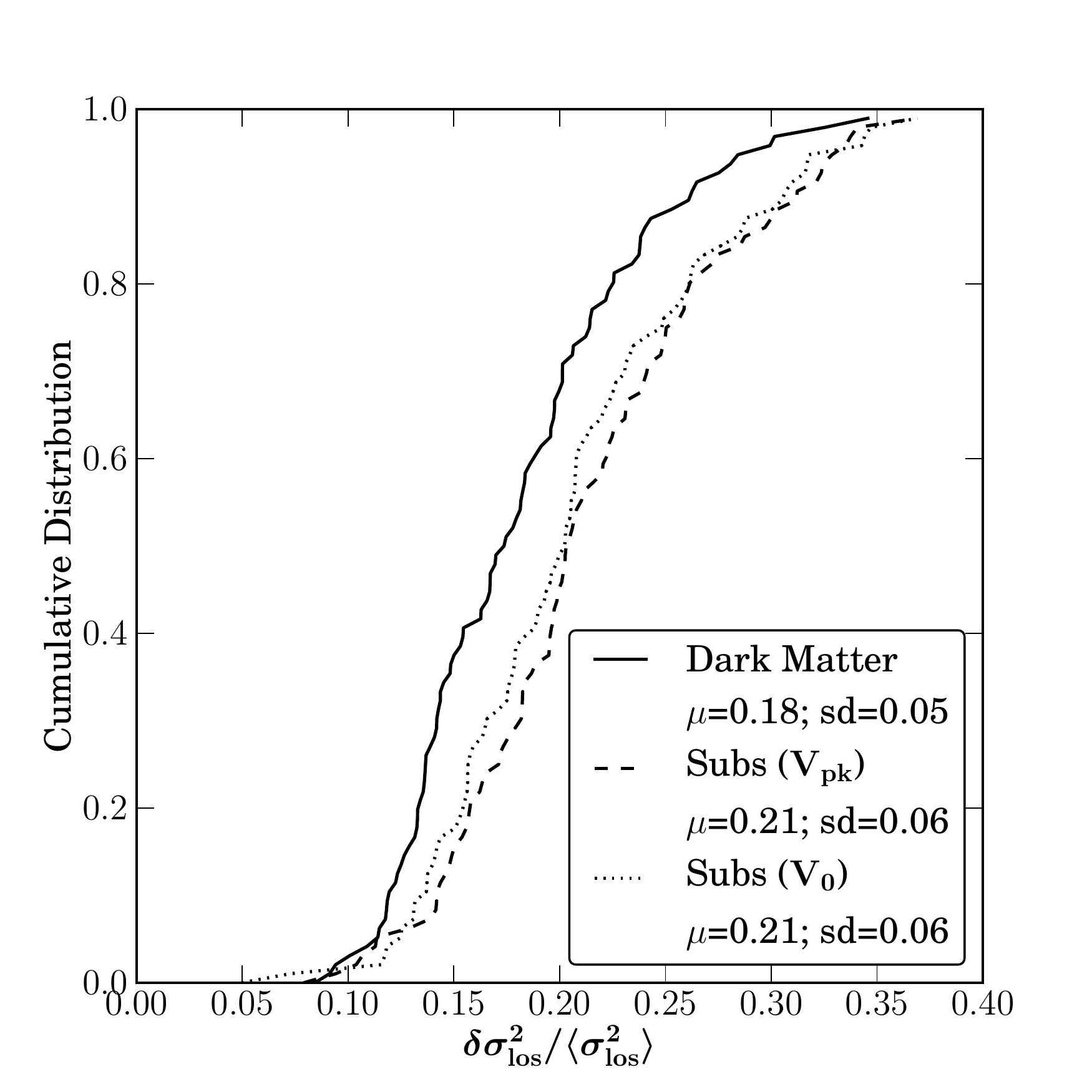}}
\caption{Distribution of the shape and velocity ellipsoid
parameters for dark matter particles and subhalos within $\Rvir$ of
the main halos.  The upper panels correspond to the shapes of spatial distribution, and
the lower panels correspond to the shapes of velocity ellipsoid.  As shown
in panels (a) and (b), dark matter particles tend to be more prolate
than subhalos.  Panel (c) shows that the velocity ellipsoids of
subhalos tend to be more elongated.  Panel (d) shows that the
statistical error of line-of-sight velocity dispersion measurements is
larger for subhalos than for dark matter particles.  }
\label{fig:shape}
\end{figure*}
%%%%%%%%%%%%%%%%%%%%%%%

In Paper I, we have discussed the shape and velocity ellipsoid of dark
matter particles of the main halo. It is interesting to see how
closely the subhalos follow the distribution of dark matter in
position and velocity space, where differences exist, and how these
depend on the specific selection of subhalos. We thus present
analogous measurements for subhalos, which are selected with $\vtoday$
and $\vpeak$.

The shape parameters are defined through the distribution tensor:
\beq
I_{ij} = \left\langle (r_i-\avg{r_i})( r_j-\avg{r_j}) \right\rangle \ ,
\eeq
where $r_i$ is the $i$th component of the position vector $r$ of a
subhalo.  The eigenvalues of $I_{ij}$ are sorted as $\lambda_1 >
\lambda_2 > \lambda_3$, and the shape parameters are defined as:
$a=\sqrt{\lambda_1}$, $b=\sqrt{\lambda_2}$, $c=\sqrt{\lambda_3}$.  We
present the dimensionless ratios $b/a$ and $c/a$.  In addition, the
triaxiality parameter is defined as
\beq
T = \frac{a^2-b^2}{a^2-c^2} \ .
\eeq
$T\approx 1$ ($a > b \approx c$) indicates a {\em prolate} halo, while
$T\approx 0$ ($a \approx b > c$) indicates an {\em oblate} halo.
Intermediate values of $T$ correspond to triaxial halos.

Analogously, the velocity ellipsoid is defined as \citep[e.g.,][]{WhiteM10}:
\beq
\sigma^2_{ij} = \left\langle (v_i-\avg{v_i})(v_j-\avg{v_j}) \right\rangle \ ,
\eeq
where $v_i$ is the $i$th component of the velocity vector.  Sorting
the eigenvalues of the velocity ellipsoid as $\lambda_1 > \lambda_2>
\lambda_3$, one can again define $a^{(v)}=\sqrt{\lambda_1}$,
$b^{(v)}=\sqrt{\lambda_2}$, $c^{(v)}=\sqrt{\lambda_3}$.

The scatter of the velocity dispersion along different lines of sight
can be calculated from the eigenvalues of the velocity ellipsoid
tensor as
\beqa
\langle \sigma^2_{\rm los} \rangle &=& \frac{1}{3}(\lambda_1+\lambda_2+\lambda_3) \\
(\delta \sigma^2_{\rm los})^2 &=& \frac{4}{45} (\lambda_1^2+\lambda_2^2+\lambda_3^2-\lambda_1\lambda_2-\lambda_2\lambda_3-\lambda_3\lambda_1) \ . \ \ \ \ \ \ \
\label{eq:sig_los}
\eeqa
We use subhalos within $\Rvir$ selected above a given threshold
without weighting them by mass in this calculation.

Figure~\ref{fig:shape} shows the cumulative distribution function of
the shape and velocity ellipsoid parameters of dark matter particles
(solid), as well as of subhalos selected with $\vtoday > 100 \ \kms$
(dotted) and subhalos selected with $\vpeak > 150\ \kms$ (dashed).
These two thresholds correspond to approximately the same number of
subhalos.  The upper panels show the axis ratios and triaxiality
parameters.  Subhalos selected by $\vtoday$ have a distribution closer
to spherical than both $\vpeak$-selected subhalos and dark matter
particles.  In addition, the dark matter distribution tends to be more
prolate than the subhalo distribution.

The lower panels of Figure~\ref{fig:shape} show the shape parameters
of the velocity ellipsoid and the line-of-sight scatter of the
velocity dispersions.  We find that the velocities of subhalos tend to
be more anisotropic than those of the dark matter particles (based on
$c^{(v)}/a^{(v)}$).  In addition, $\delta \sigma^2_{\rm los}$ is higher
for subhalos than for dark matter particles.  This trend can be
understood based on the findings of \cite{WhiteM10}: the motions of
subhalos tend to be anisotropic because they can retain their infall
velocities for a long time.  In contrast, the merged and stripped
material is dynamically older and contributes to a well-mixed
isotropic velocity distribution.  We also note that because there are
many more dark matter particles than subhalos, these coherent subhalos
contribute significantly to the velocity ellipsoids of subhalos but
negligibly to the velocity ellipsoids of dark matter particles.
However, we find that the difference in $c^{(v)}/a^{(v)}$ between
subhalos and dark matter does not correlate with any of the formation
time proxies, indicating that formation history may not fully account
for this difference.

We summarize several trends in Figure~\ref{fig:shape} and propose explanations:
\begin{itemize}

\item The distribution of subhalos tends to be more spherical than
that of dark matter.  This can be understood by the fact that subhalos
have a shorter relaxation time than dark matter particles, because
$t_{\rm relax} \approx (R/v)(N/\ln N)$ and the $N$ for subhalos is much
smaller.

\item The velocity ellipsoid of subhalos tends to be more elliptical
than that of dark matter.  This can be explained by the anisotropic
motions of subhalos accreted as a group (as discussed above).

\item Subhalos selected with $\vtoday$ tend to have a more spherical
distribution than those selected with $\vpeak$.  Similar trend exists
for $c^{(v)}/a^{(v)}$ but does not exist for $b^{(v)}/a^{(v)}$.  This
trend can be understood through stripping: a $\vpeak$ selection tends
to include more highly stripped subhalos than $\vtoday$, and
highly stripped subhalos tend to be on more elliptical orbits
(stripping will be stronger for those orbits that have a smaller
pericenter), leading to the higher ellipticity measured with $\vpeak$.
In addition, we note that overmerging will also make the distribution
of subhalos more spherical, because overmerging tends to eliminate
highly stripped subhalos, which tend to have more elliptical orbits
(again due to the smaller pericentric distance).

\end{itemize}

We note that for both $\vtoday$ and $\vpeak$ selection, when we
increase the threshold, the ellipticity slightly increases.  This
trend can be explained by the statistical biases arising when a
smaller number of subhalos is used to measure the ellipsoids.  To
confirm this, we randomly select a number of subhalos within $\Rvir$
with $\avg{N}$ matching either the number of subhalos obtained with
the $\vtoday$- or $\vpeak$-selection, and we recover the trend that a
smaller number of subhalos always leads to a higher inferred
ellipticity.  Although it is possible that large radius is weighted
more in the position tensor and that small radius is weighted more in
the velocity tensor, we find that subhalos of different masses do not
have radial distributions that are distinct enough to explain the
trend with selection threshold.

In addition, we observe that the difference between shapes measured by
dark matter particles and subhalos has a slight trend with the
formation history --- for halos that experienced recent major mergers,
the subhalo distribution tends to be much rounder than the total dark
matter particle distribution.  This could also be related to the fact
that subhalos have a shorter relaxation time than dark matter
particles.  However, we note that this trend with formation history is
rather weak and has a large scatter, indicating that formation time
and relaxation cannot fully explain the shapes.

Finally, we note that in all cases, the differences in axis ratios are
at the level of a few percent.  Observationally, these differences are
likely to be overwhelmed by the scatter due to line-of-sight
projection, viewing angle, spectroscopic sample selection, etc.  These
effects are likely to depend on the environment and details of the
observational techniques employed.  It would be interesting to
investigate whether and, if so, how the difference between the
velocity ellipsoids for the various selection criteria depends on
environment \citep[see also][who have demonstrated a dependence of
subhalo kinematics on environment in the Millennium
Simulation]{Faltenbacher2010}.  An analysis of the environmental
dependence of halo properties will be deferred to a future paper.

\section{Correlation between subhalo properties and formation history}\label{sec:cor}

In Paper I, we have focused on the impact of the formation history on
the density profile and the phase-space structure of the cluster
halos. Here, we investigate the impact of the formation history on the
subhalo population.  In Figure~\ref{fig:correlation_sub}, we present
the correlation of eight quantities measured from our sample: four for
subhalo properties, three for formation time, and one for the halo
profile.

\subsection{Formation History Parameters}

As discussed in Paper I, to quantify the formation history of halos,
we adopt an exponential-plus-power law model with two parameters
\citep{McBride09}
\beqa
M(z) &=& \Mtoday (1+z)^\beta e^{-\gamma z} \ ,\\
-\frac{d\ln M}{dz} &\approx&  \gamma-\beta \ \ \mbox{when $z\ll1$}\ .
\eeqa
Thus, $\gamma-\beta$ provides a measure of the late-time accretion rate.

We use $\zhalf$, the earliest redshift that a halo obtains half of its
mass, as the formation time proxy throughout the paper, but note that
using formation time proxies based on fitting functions (as those
studied in Paper I) lead to similar results.  In addition, we include
the redshift of the last major merger of each main halo, $z_{\rm lmm}$,
defined as the last time a halo with a mass ratio of at least 1/3
crossed the virial radius and became a subhalo.

We note that the formation history is directly reflected in the halo
density profile, which has been studied in detail in Paper I.  For the
completeness of our correlation analysis, we also include the halo
concentration defined for the Navarro--Frenk--White (NFW) profile
\citep{NFW97}:
\beq
\frac{\rho(r)}{\rho_{\rm crit}} = \frac{\delta_c}{(r/r_s)(1+r/r_s)^2} \ ,
\eeq
for which the concentration parameter is defined as 
\beq
c_{\rm NFW} = \frac{\Rvir}{r_s} \ .
\eeq
We kindly refer the reader to Paper I for details on the fitting procedure.

\subsection{Subhalo Mass Fraction}
%%%%%%%%%%%%%%%%%%%%%%%
\begin{figure*}
\hspace*{-0.8in}
\includegraphics[width=2.5\columnwidth]{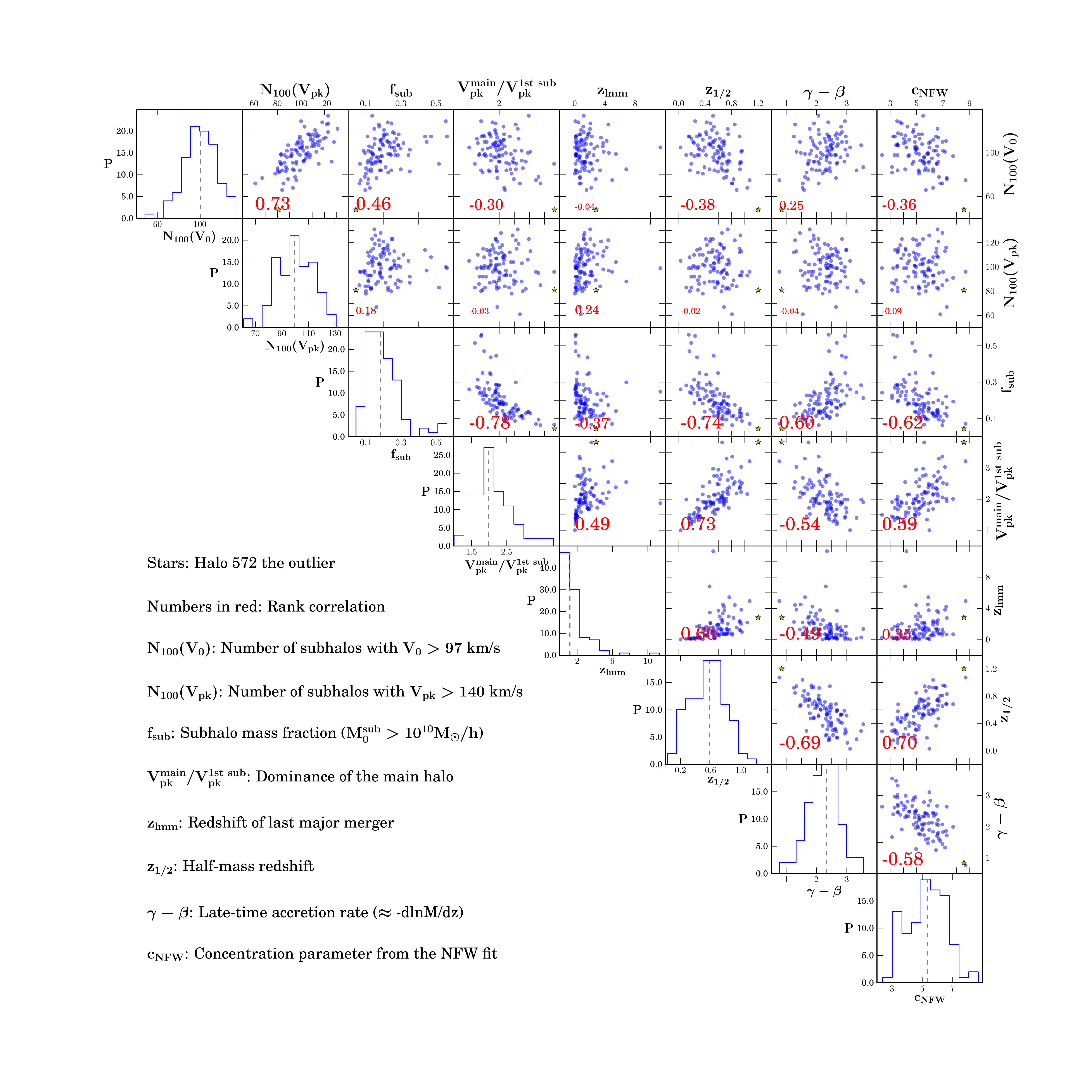}
\caption[]{Correlation between the subhalo properties and the
formation time proxies. The top two rows show the number of subhalos
selected with $\vtoday$ and $\vpeak$.  In each case, the threshold is
chosen to have an average of 100 subhalos per main halo.  Correlations tend
to be weaker with subhalos selected by $\vpeak$ than by $\vtoday$,
except for $z_{\rm lmm}$.}
\label{fig:correlation_sub}
\end{figure*}
%%%%%%%%%%%%%%%%%%%%%%%

We next investigate the mass of the main halos that is contained in subhalos.
The subhalo mass fraction can be defined as
\beq
f_{\rm sub}(M_{\rm th}) = \frac{1}{M_{\rm main}}\sum_{M_{\rm sub}> M_{\rm th}} M_{\rm sub} \ 
\eeq
and can be used as an indication of a recent major merger event.  For
example, if a massive subhalo accreted onto the main halo only
recently, it retains most of its mass and contributes to a large
$f_{\rm sub}$.  We show below that $f_{\rm sub}$ is correlated with halo
formation time and can thus be used as an indicator of the state of
relaxedness of the halo \citep[e.g.,][]{DeLucia04, Shaw06}.  Here we
choose $M_{\rm th} = 10^{10}\hiMsun$, which approximately corresponds to
our completeness limit (see Figure~\ref{fig:SHMF}), but note that the
correlations presented below are insensitive to the specific choice of
$M_{\rm th}$.

The third row and column of Figure~\ref{fig:correlation_sub}
correspond to $f_{\rm sub}$ and show, as expected, that $f_{\rm sub}$ is
strongly correlated with $z_{1/2}$, $\gamma-\beta$, and $c_{\rm NFW}$.
That is, halos of higher $f_{\rm sub}$ tend to be late forming, with high
late-time accretion rates and associated low concentration.

The subhalo mass fraction itself is also a quantity of observational
interest.  For example, it can be inferred from gravitational lensing
\citep[e.g.,][]{DalalKochanek02, Vegetti12, Fadley12}.  Accurate
modeling of the subhalo mass fraction is essential for the study of
the lensing flux ratios \citep[e.g.,][]{Xu09}.  For the relatively
massive systems considered here, we find that $f_{\rm sub}$ is strongly
correlated with the mass of the most massive subhalo, despite the fact
that this halo contributes on average only $\sim 20\%$ of the total
subhalo fraction.  We also find that the subhalo fraction is strongly
correlated with the dominance of the main halo (related to the
luminosity gap between brightest and second brightest galaxies), as
well as with formation time and concentration.  Because strong lensing
clusters tend to have higher than average halo concentrations, it is
important to take this correlation into account when interpreting
measurements of the subhalo mass fraction from strong lensing.

\subsection{Mass Contributed by Merged Subhalos}

%%%%%%%%%%%%%%%%%%%%%%%%%%%%%%%%%%%
\begin{figure}
\centering
\includegraphics[width=\columnwidth]{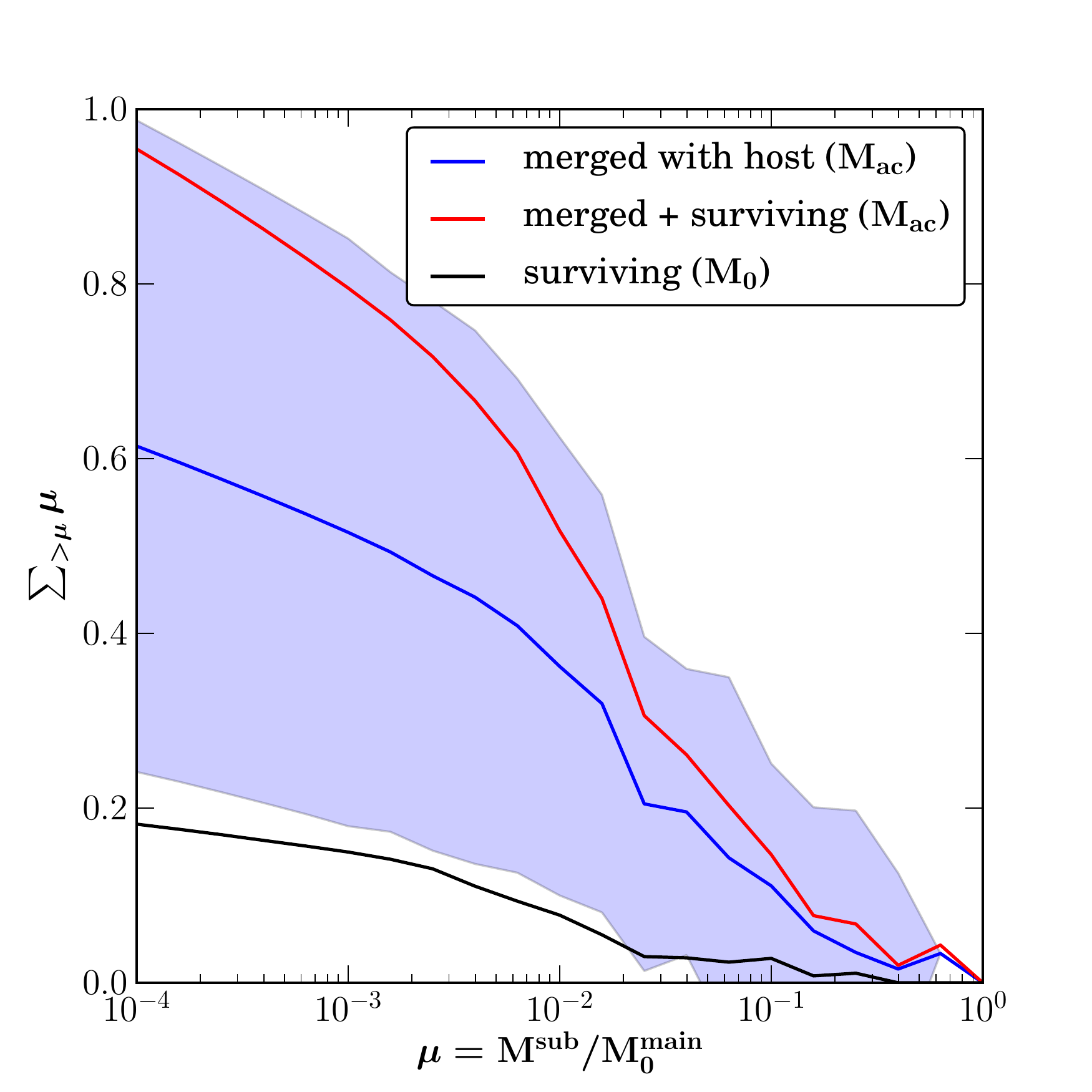}
\caption[]{Contribution of mass to the main halo from subhalos
above a certain mass ratio $\mu = M^{\rm sub}/M_0^{\rm main}$.  The blue curve
corresponds to the contribution from subhalos that have merged with
the main halo, and we use the subhalo mass at accretion, $\Macc$; the
region enclosed by blue dotted curves
corresponds to the 68 \% scatter for the sample.  The
red curve includes merged subhalos and those that are surviving, for
which we also use $\Macc$.  The black curve corresponds to subhalos that
are still surviving today, for which $\Mtoday$ is used, and is
equivalent to $f_{\rm sub}(>\mu)$.}
\label{fig:merging}
\end{figure}
%%%%%%%%%%%%%%%%%%%%%%%%%%%%%%%%%%%

In the previous subsection, we addressed the mass contributed by the
{\em present} subhalo population.  We now investigate the mass that
was brought into the main halo by all merging events in a halo's
history.  Figure~\ref{fig:merging} shows the contribution to the main
halo mass from merged subhalos.  This has also been explored by, e.g.,
\cite{Berrier09} for lower mass systems.  The $x$-axis corresponds to
the ratio of subhalo mass to main halo mass, $\mu =
M^{\rm sub}/M_{0}^{\rm main}$, and the $y$-axis corresponds to the fraction of
main halo mass contributed by subhalos above a given $\mu$.

Here we consider two types of subhalos.  The {\em first type} is those
subhalos that have merged into the main halo and can no longer be
identified; for this type of subhalo, we use its mass when it accreted
onto the main halo, $\Macc$.  These subhalos are represented by the
blue curve, and the region enclosed by blue dotted curves corresponds
to the 68\% scatter about this contribution for the sample.  Our
results indicate that, on average, 60\% of the main halo's mass comes
from merged subhalos with $\mu > 10^{-4}$; however, this number varies
greatly from halo to halo.

The {\em second type} is those subhalos that still survive today
(i.e., that can be identified by the halo finder at $z=0$).  For this
type, we also use $\Macc$. Since subhalos tend to lose a significant
amount of mass due to tidal stripping inside the main halos, using
$\Macc$ ensures that we include the mass that once belonged to
subhalos but later got stripped by the main halo.  We do not
explicitly count the subhalos that merge into other subhalos because
their masses have already been included in the surviving subhalos.
The red curve corresponds to the sum of the first and the second
types. We find that more than 90\% of the mass can be attributed to
halos with $\mu > 10^{-4}$ that were accreted onto the main halo.

Finally, the black curve shows the contribution to main halo from
subhalos that survive at $z=0$.  We use the current mass of these
subhalos, $\Mtoday$; this quantity is equivalent to the subhalo mass
fraction for different mass thresholds, $f_{\rm sub}(>\mu)$.  Subhalos
with $\mu > 10^{-4}$ constitute a little less than $20\%$ of the mass
of the main halo.

\subsection{Dominance of the Main Halo}

In this section, we study the difference between each main halo and
its largest subhalo, a quantity motivated by the definition of the
so-called fossil groups.  Observationally, these systems are
defined as having a large magnitude gap between the brightest and
second brightest galaxies, in addition to being X-ray luminous
\citep[e.g.,][]{TremaineRichstone77,Jones03,Miller12}.  Fossil systems
are often interpreted as a population of galaxy groups that have
assembled at early times and have not undergone a recent major merger.

Since predicting the optical and X-ray properties of our halos is
beyond the scope of this paper, we define a related property, the
ratio of $\vpeak$ for the main and the first subhalo (the subhalo
with highest $\vpeak$):
\beq
D = \frac{\vpeak^{\rm main}}{\vpeak^{\rm 1st\ sub}}.
\eeq
We note that using the second subhalo leads to the same trend presented below.

In Figure~\ref{fig:correlation_sub}, the fourth column and row show
that $D$ is correlated with the formation time, late-time accretion,
concentration, and subhalo mass fraction.  We also note that all our
main halos have similar $\vpeak^{\rm main}$; therefore, the scatter in
$D$ is almost completely determined by $\vpeak^{\rm 1st\ sub}$.

The trends observed in Figure~\ref{fig:correlation_sub} can be
understood as follows. Since $\vpeak^{\rm 1st\ sub}$ indicates the
maximum of the subhalo mass that accretes onto a main halo, a main
halo with a low $\vpeak^{\rm 1st\ sub}$ has fewer massive subhalos
accreting onto it (this is also reflected by its low $f_{\rm sub}$).  With
relatively fewer incoming subhalos, to achieve the same mass today,
these halos must have obtained most of their mass at early times and
have undergone slow accretion at late time, thus leading to the low
$\gamma-\beta$ and the high concentration.

While we were preparing this manuscript, we learned about the related
work of \cite{Hearin12}, who have studied the ``magnitude gap'' of the
two brightest cluster members, which is analogous to our dominance
parameter $D$.  These authors have found that for SDSS groups of a
given velocity dispersion, clusters with high magnitude gap tend to
have low richness, and this correlation can in turn reduce the scatter
in the mass inferences using optical mass tracers.  In contrast, we
find that the number of subhalos selected with $\vpeak$ is not
correlated with $D$ for halos of the same mass.  Since the results
from \cite{Hearin12} are based on velocity dispersion rather than
mass, a fair comparison between our results and theirs will require
further consideration of the scatter in velocity dispersion, scatter
of galaxy luminosity at a given $\vpeak$, as well as observational
selections, which are beyond the scope of the current work.

\subsection{The Curious Case of Halo~572: An Outlier and a Fossil Cluster}
\label{sec:572}

In the {\sc Rhapsody} sample, we find one peculiar halo---Halo
572---which is a prominent outlier in formation time (highest
$\zhalf$) and occupies the tail of many halo properties as well as the
corner of several scatter plots (marked as stars in
Figure~\ref{fig:correlation_sub}).  It has unusually high $c_{\rm
NFW}$ (2.7$\sigma$ deviation from the mean) and central dominance
(3.2$\sigma$). It also has one of the lowest late-time accretion rates
$\gamma-\beta$ (2.7$\sigma$), $f_{\rm sub}$ (1.6$\sigma$), and subhalo
numbers selected with several different criteria.  This halo obtained
most of its mass at early time and nearly stopped accreting mass at
late time, leading to these extreme properties.  Images of the
evolution of Halo~572 are shown in Figure~3 of Paper~I, where it is
evident that this halo had an atypical formation history.  We find
that Halo~572 does {\em not} live in an atypical environment on large
scales.

The high central dominance indicates that, if such a halo is observed,
it would likely have a large luminosity gap between the brightest and
second brightest galaxies, and its high concentration will make it
X-ray luminous.  Therefore, we expect that this halo will host a
cluster that satisfies the criteria of a ``fossil.''  In addition, it
is a ``real'' fossil cluster in the sense that it has an unusually
early formation history.  Studies of fossil groups in both simulations
and observations have come to a range of conclusions, with debate
about whether fossil groups have distinct assembly histories or are
merely an intermediate state in galaxy formation (see, e.g.,
\citealt{Cui11,Sales07}, and references therein).  Halo~572 presents a
case of a distinctively early formation history and the consequential
properties.  From these results we conclude that it is highly probable
that ``real fossils'' exist in the universe but that they are very
rare; thus, they require more stringent selection criteria to be
distinguished from some transient states of cluster formation.

\section{Correlation between formation time and subhalo number:
The impact of subhalo selection}\label{sec:nocor}

%%%%%%%%%%%%%%%%%%%%%%%%%%%%%%%%%%%%%%%%%%%%%%%%%%%%%%%%%%%%
\begin{figure*}
\centering
\includegraphics[width=\columnwidth]{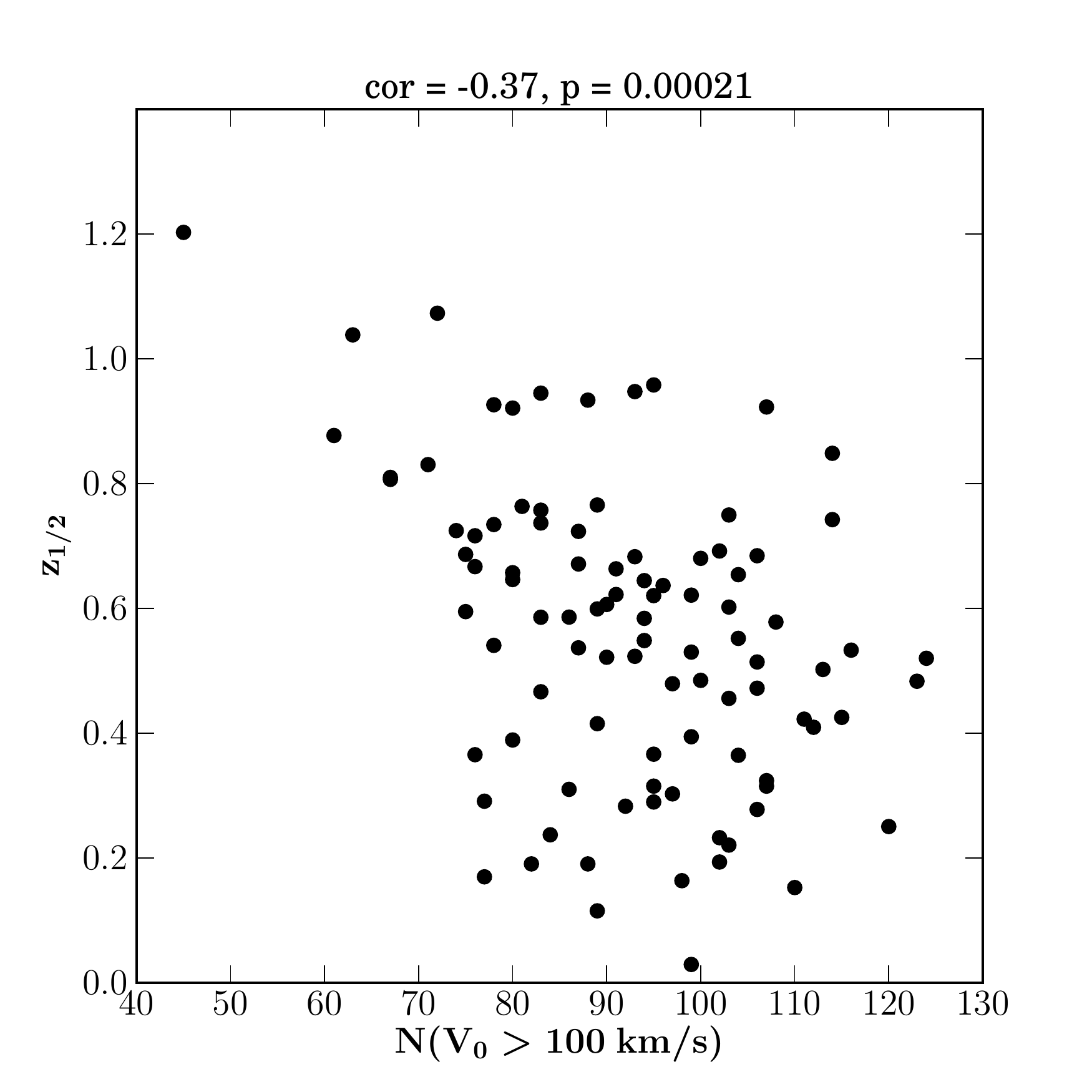}
\includegraphics[width=\columnwidth]{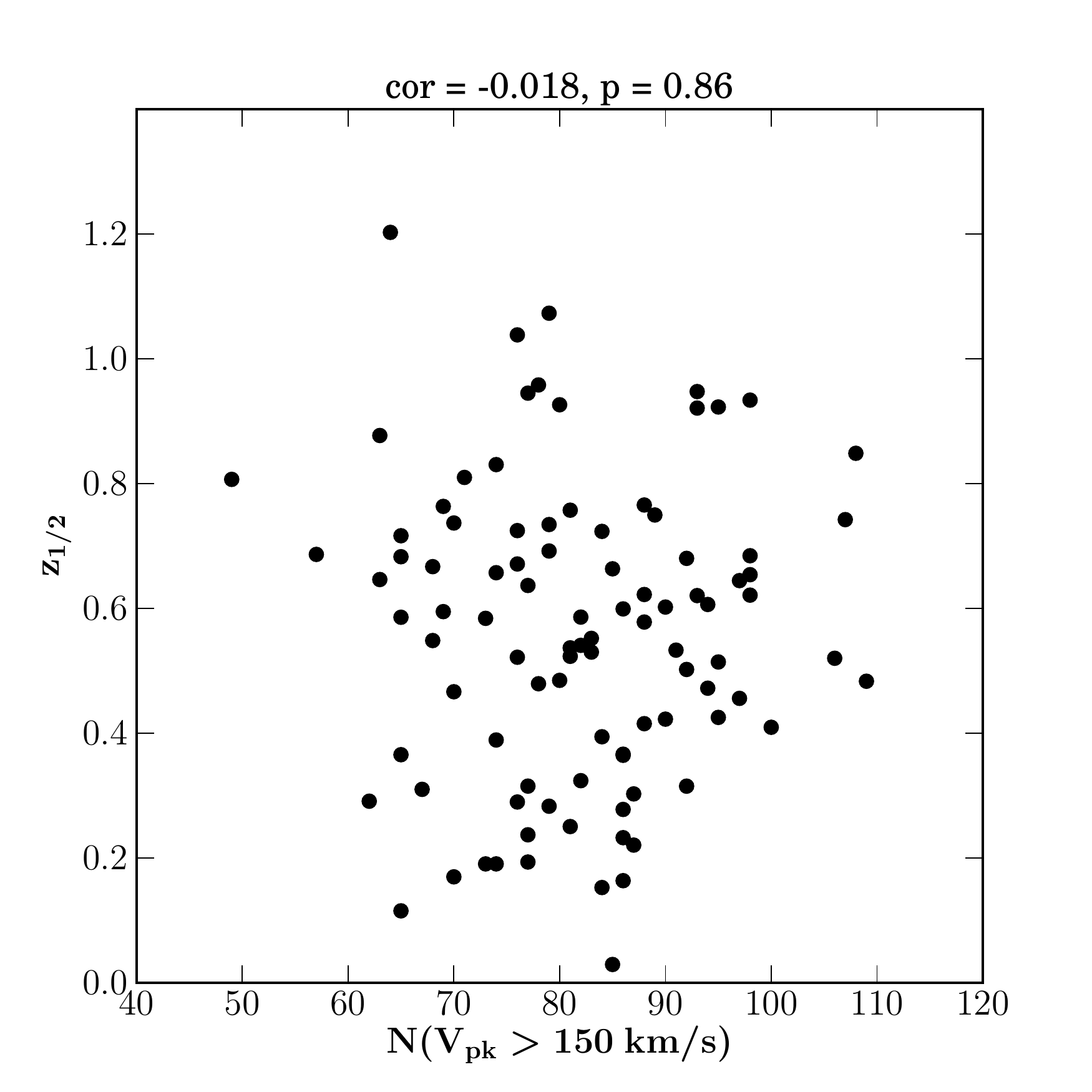}
\caption[]{Impact of subhalo selection on the correlation between
subhalo number and formation time $\zhalf$.  The left panel
corresponds to selecting subhalos using $\vtoday > 100\ \kms$, while
the right panel corresponds to $\vpeak> 150\ \kms$.  Although the
$\vtoday$ selection presents significant anti-correlation between
subhalo number and formation time, the $\vpeak$ selection presents no
such correlation.  This trend can be explained by the stripping of
subhalos, as demonstrated by the following two figures.  We note that
the Spearman rank correlation coefficient and the $p$-value are quoted
on each panel.}
\label{fig:N_zhalf_correlation}
\end{figure*}
%%%%%%%%%%%%%%%%%%%%%%%%%%%%%%%%%%%%%%%%%%%%%%%%%%%%%%%%%%%%
\begin{figure*}
\centering
\includegraphics[width=\columnwidth]{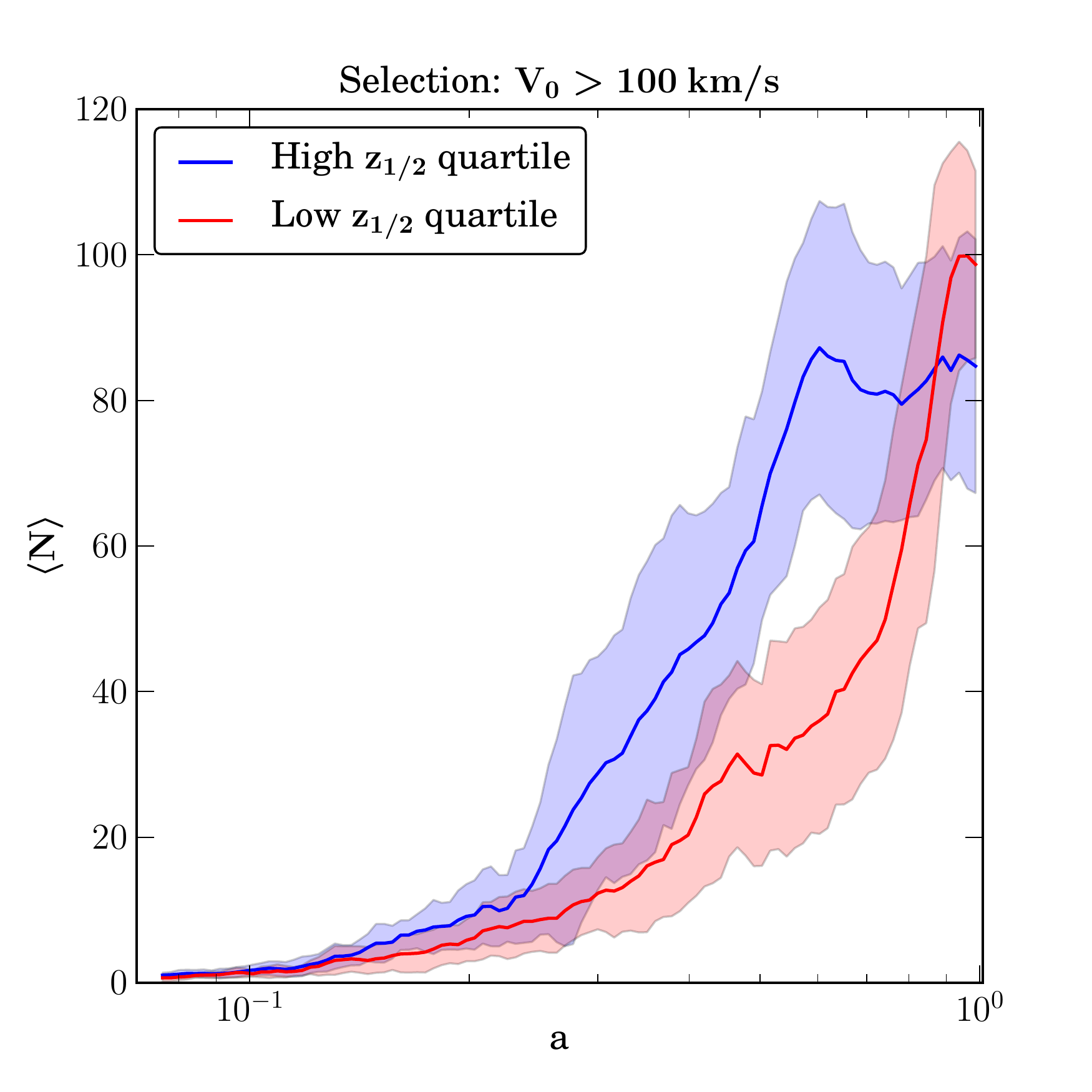}
\includegraphics[width=\columnwidth]{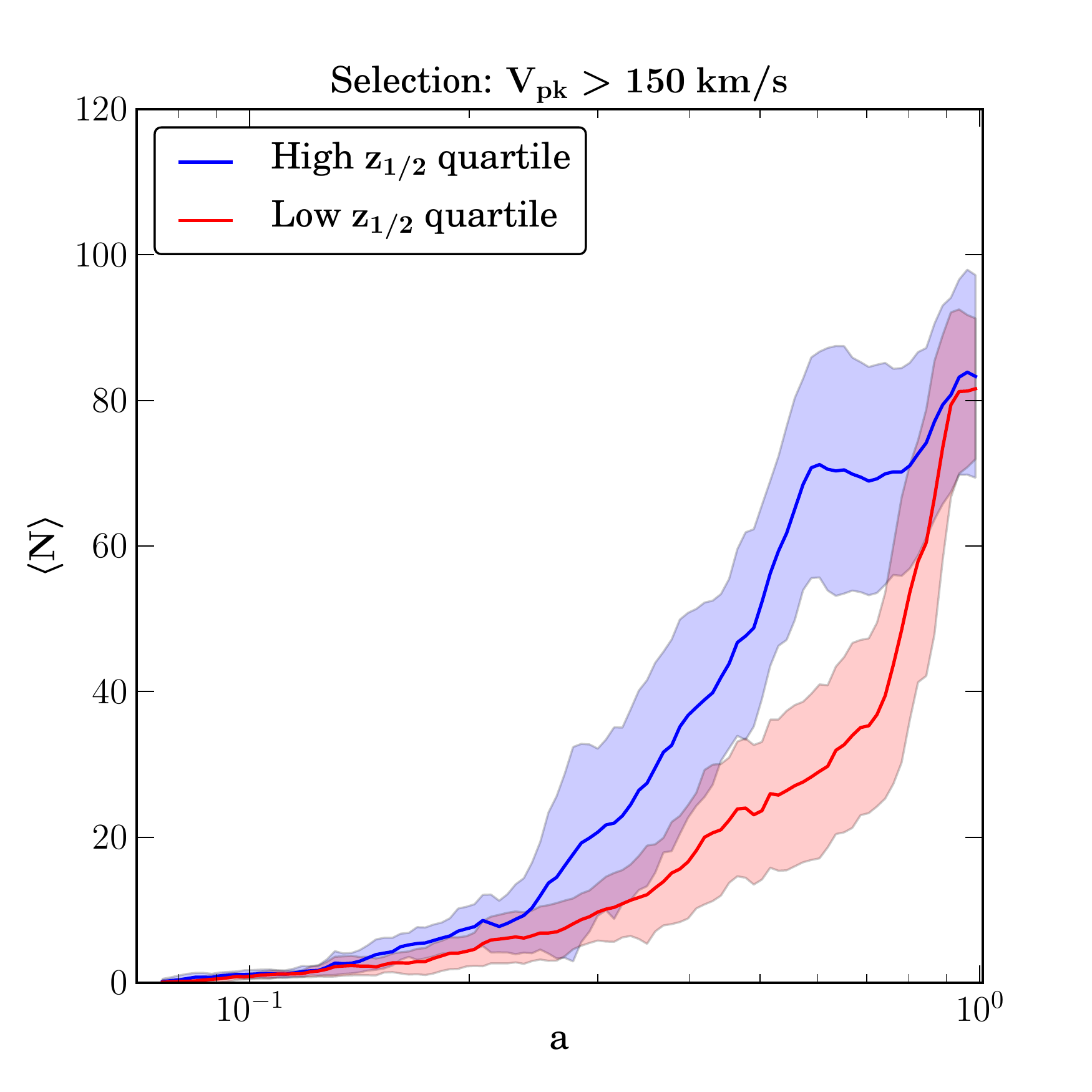}
\caption[]{Evolution of subhalo number, split by quartiles of
$\zhalf$.  Left: subhalos with $\vtoday > 100 \ \kms$; right: subhalos
with $\vpeak > 150 \kms$.  When subhalos are selected with $\vtoday$,
the subhalo number of early-forming (blue) and late-forming (red)
halos split at $z=0$; however, when subhalos are selected with
$\vpeak$, there is no clear split of halo number at $z=0$.  This trend
is reflected by the difference in the correlation seen in
Figure~\ref{fig:N_zhalf_correlation}.}
\label{fig:SAH_zhalf}
\end{figure*}
%%%%%%%%%%%%%%%%%%%%%%%%%%%%%%%%%%%%%%%%%%%%%%%%%%%%%%%%%%%%
\begin{figure}
\centering
\includegraphics[width=\columnwidth]{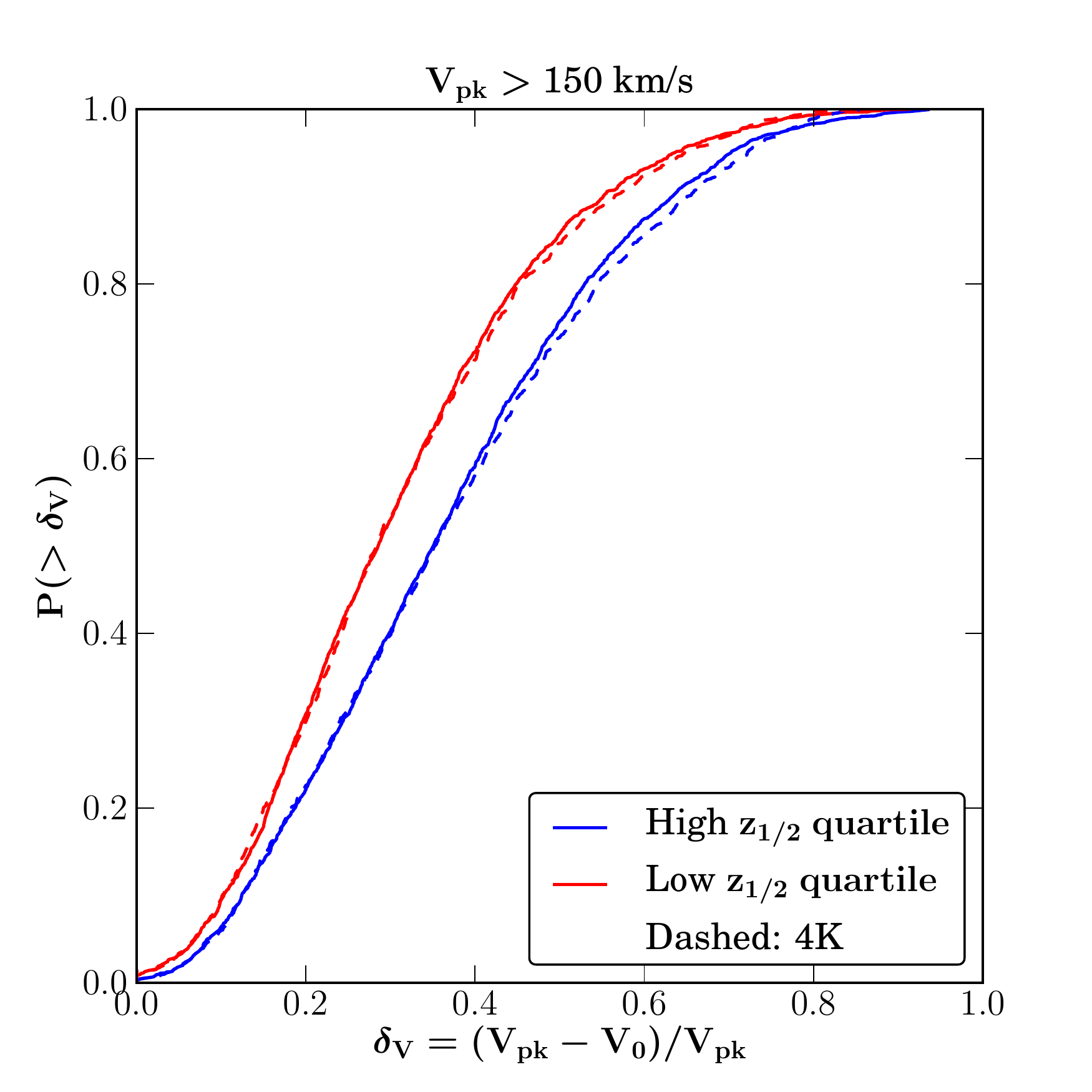}
\caption[]{Cumulative distribution function of $\delta_{\vmax} =
(\vpeak - \vtoday)/\vpeak$, an indication of the amount of stripping
experienced by subhalos.  We split the main halo by the formation time
$\zhalf$.  Early-forming halos (blue) tend to have subhalos with
higher $\delta_{\vmax}$ (stronger stripping) on average than
late-forming ones (red).  If subhalos are selected with $\vtoday$,
highly stripped subhalos tend to fall below the threshold, leading to
a low subhalo number.  This can explain the correlation seen in the
$\vtoday$ selection in Figures~\ref{fig:N_zhalf_correlation}
and~\ref{fig:SAH_zhalf}.}
\label{fig:delta_vmax}
\end{figure}
%%%%%%%%%%%%%%%%%%%%%%%%%%%%%%%%%%%%%%%%%%%%%%%%%%%%%%%%%%%%
\begin{figure}
\includegraphics[width=\columnwidth]{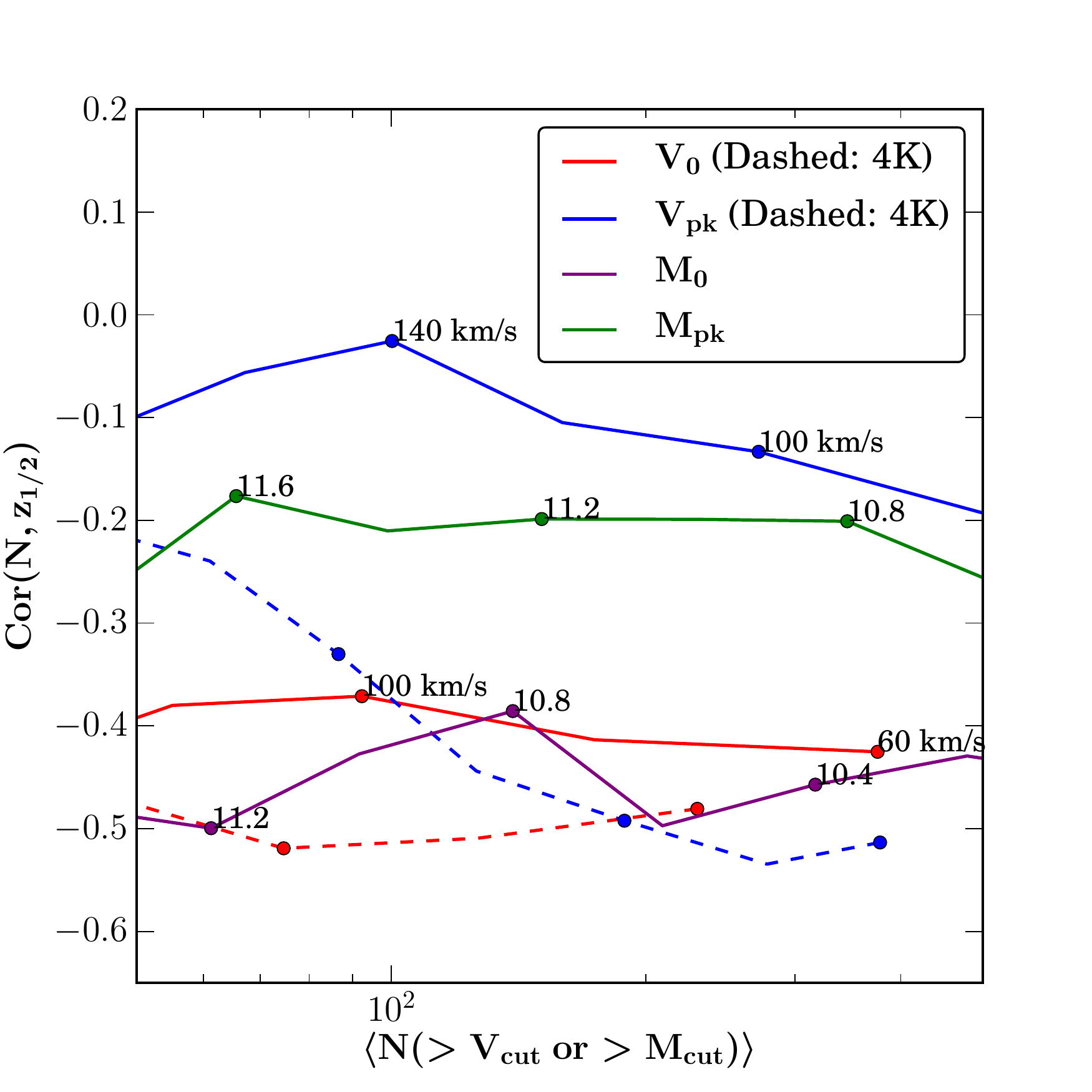}
\centering
\caption{Correlation between subhalo number and $\zhalf$, for
different subhalo selection methods.  When subhalos are selected with
$\vtoday$ or $\Mtoday$ (red and purple), an anti-correlation exists
for all thresholds; when subhalos are selected with $\vpeak$ or
$\Mpeak$ (blue and green), the anti-correlation is greatly reduced or
non-existent.  In addition, the comparison between 8K and 4K (solid
and dashed of the same color) sample shows that the anti-correlation
can be enhanced by insufficient resolution.}
\label{fig:vcut_Mcut_correlation}
\end{figure}
%%%%%%%%%%%%%%%%%%%%%%%%%%%%%%%%%%%%%%%%%%%%%%%%%%%%%%%%%%%%

In this section, we investigate in detail the correlation between
formation time and subhalo abundance (as previously explored by e.g.,
\citealt{Zentner05,Wechsler06,Giocoli10}).  We show that this
correlation is mainly caused by subhalo stripping and insufficient
resolution.  However, in the real universe, stripping of dark matter
particles is less relevant to the galaxy content in clusters, and this
correlation is therefore not expected to exist for observable
galaxies.

\cite{Zentner05} found that early-forming halos tend to have fewer
subhalos.  In their study, subhalos are selected with a threshold on
$\vtoday$ ($\vmax$ at $z=0$).  This trend has been explained by the
fact that in early-forming halos, subhalos tend to accrete at early
time and are more likely to be destroyed, which leads to a low number
of subhalos.  However, we find that this correlation strongly depends
on how subhalos are selected.  In what follows, we explore the
dependence of this correlation on various subhalo selection criteria
and understand the trend by investigating the accretion and stripping
of subhalos.

Figure~\ref{fig:N_zhalf_correlation} demonstrates how subhalo
selection based on $\vtoday$ or $\vpeak$ can lead to different
correlations between formation time and subhalo number.  The left
panel corresponds to selecting subhalos with $\vtoday> 100 \ \kms$,
and the right panel corresponds to $\vpeak > 150\ \kms$.  We note
these two thresholds correspond to roughly the same number of
subhalos.  Each point corresponds to a main halo in our sample.  The
$x$-axes correspond to the number of subhalos under either selection
criterion, and the $y$-axes correspond to the formation time proxy
$\zhalf$.  When subhalos are selected based on $\vtoday$ (left panel),
formation time and subhalo number are significantly anti-correlated.
However, when subhalos are selected based on $\vpeak$ (right panel),
this anti-correlation no longer exists in our sample.

This lack of anti-correlation can be understood as follows.  When
subhalos are selected with a given threshold of $\vtoday$, the
stripping of subhalos directly impacts the subhalo number: for a halo
that assembled earlier, its subhalos experience stripping for a longer
time and with a higher intensity (because of the high halo
concentration), and the subhalos' masses and $\vtoday$ tend to be
greatly reduced.  Therefore, fewer subhalos remain above the $\vtoday$
threshold.  On the other hand, when we select subhalos using $\vpeak$,
the stripping of subhalos does not directly impact the subhalo number,
as long as those subhalos are still identifiable.

To support our argument above, we investigate the evolution of the
subhalo population for these two subhalo selection criteria.
Figure~\ref{fig:SAH_zhalf} shows subhalo number as a function of the
scale factor a (we plot in $\log a$ to emphasize the late-time
behavior).  As in the previous figure, the left/right panel
corresponds to the $\vtoday$/$\vpeak$ selection.  In both panels, we
plot the subhalo number evolution for the highest $\zhalf$ quartile
(blue) and lowest $\zhalf$ quartile (red).

In the left panel of Figure~\ref{fig:SAH_zhalf}, we can see that for
early-forming halos (blue), the subhalo accretion rate is high at early
time but suddenly declines after $a\approx0.6$.  On the other hand,
for the late-forming halos (red), their subhalo accretion rate is high
at late time.  At $a\simeq1$, the early-forming halos have fewer subhalos
than late-forming halos.  In the right panel,
although the early-forming halos have declined in subhalo accretion
rate at late time, their subhalo number still grows at late
times, and their subhalo numbers are similar to late-forming halos at
$a\simeq1$.

The different trends in both panels can be attributed primarily to the
stripping of subhalos.  Figure~\ref{fig:delta_vmax} shows the
cumulative distribution of the fractional change of $\vmax$ of
subhalos
\beq
\delta_{\vmax} = \frac{\vpeak-\vtoday}{\vpeak} \ .
\eeq
This quantity can be used as a measure of the amount of stripping
experienced by subhalos.  Higher $\delta_{\vmax}$ indicates that a
subhalo has experienced stronger or longer stripping and has lost more
mass.  For early-forming halos (blue), subhalos on average have higher
$\delta_{\vmax}$, indicating that these subhalos experience more
stripping and their $\vmax$ is reduced more.  As a result, if we
select subhalos using $\vtoday$, we tend to exclude subhalos that have
experienced more stripping.  These subhalos will however be included
if we select subhalos using $\vpeak$.

Therefore, the correlation between formation time and subhalo number
seen in a selection on $\vtoday$ can be attributed to the exclusion of
highly stripped subhalos.  Since subhalos selected with $\vtoday$ have
less observational relevance than those selected with $\vpeak$, our
results imply that cluster richness is unlikely to be correlated with
the formation time of the halo in observations.

So far we have been using two specific selection thresholds for
$\vtoday$ and $\vpeak$.  Here, we investigate how our results depend
on the selection threshold.  In
Figure~\ref{fig:vcut_Mcut_correlation}, we present the correlation
between subhalo number and halo formation redshift, $Cor(N, \zhalf)$,
where $N$ is the subhalo number above some selection thresholds.  We
discuss four selection methods: $\vtoday$, $\vpeak$, $\Mtoday$, and
$\Mpeak$.  For each selection threshold, we compute the mean number of
subhalos, $\avg{N}$.  Using $\avg{N}$ as the $x$-axis allows us to put
these curves on the same figure.

In Figure~\ref{fig:vcut_Mcut_correlation}, the different magnitudes of
correlation can easily be seen.  Here, we compare four pairs of
selection methods:
\begin{itemize}

\item {\em $\vtoday$ versus $\vpeak$ (red versus blue).} The former
has stronger correlation with $\zhalf$ due to subhalo stripping, as
discussed above.

\item {\em $\Mtoday$ versus $\Mpeak$ (purple versus green).} The
former has stronger correlation, for the same reason as above.

\item {\em $\vtoday$ versus $\vtoday$ $4K$ (red solid versus red
dashed).} The latter has stronger correlation, indicating that an
unphysical correlation can be introduced by insufficient resolution.

\item {\em $\vpeak$ versus $\vpeak$ $4K$ (blue solid versus blue
dashed).} The latter has stronger correlation, indicating that using
$\vpeak$ does not mitigate the impact of resolution.  For other
quantities, comparisons between 8K and 4K show the same trend.

\end{itemize}

Since $\vpeak$ is more relevant for observations than $\vtoday$,
$\Mtoday$, and $\Mpeak$, the lack of correlation when selecting by
$\vpeak$ indicates that, observationally, the formation time of a
galaxy cluster is unlikely to be inferred from the number of galaxies
alone.  Thus, richness-selected galaxy clusters are unlikely to be
biased in terms of their formation time, implying that the effect of
assembly bias may be negligible for cluster cosmology self-calibration
\citep{WuHY08}.

We now return to the discussion of Figure~\ref{fig:correlation_sub}.
We have shown that various observables, including the subhalo
fraction, the central dominance, and the concentration, are highly
correlated with formation time and the amount of late-time accretion.
These observables are also correlated with the number of subhalos
selected by the current maximum circular velocity, $\vtoday$.
However, these correlations largely disappear when selecting subhalos
based on $\vpeak$, which is expected to be more strongly correlated
with galaxy luminosity or stellar mass \citep{Conroy06, Reddick12}.
This reduces the likelihood that these observables provide additional
mass information for richness-selected samples of galaxy clusters.

%%%%%%%%%%%%%%%%%%%%%%%%%%%%%%%%%%%%%%%%%%%%%%%%%%%%%%%%
\section{Summary and Discussion}\label{sec:summary}

In this paper, we have presented the key properties of the subhalo
populations in the {\sc Rhapsody} cluster re-simulation project, a
sample of 96 halos of $\Mvir=10^{14.8\pm 0.05}\hiMsun$, resolved with
approximately $5\times10^6$ particles inside the virial radius.  We
focus on the effect of formation history on the subhalo population.
We find that this effect depends on subhalo selection criteria and
resolution, which need to be carefully taken into account to make
observationally relevant inferences. Our findings can be summarized as
follows:

\begin{enumerate}

\item {\em Subhalo statistics.} In Section~\ref{sec:SHMF}, we show the
subhalo mass function for several subhalo mass proxies: $\vtoday$,
$\vpeak$, and $\Mtoday$.  We find that for a given halo, the numbers
of large and small subhalos are only moderately correlated with each
other.  In Section~\ref{sec:Poisson}, we compare the scatter in
subhalo number under different selection criteria and resolutions,
finding that subhalo stripping and insufficient resolution can lead to
extra non-Poisson scatter.  The least stripped proxy, $\vpeak$ (8K),
still has a small amount of residual scatter above Poisson statistics,
corresponding to a constant value of $\alpha$ = 1.005.

\item {\em Shape of spatial distribution and velocity ellipsoid.} In
Section~\ref{sec:sub_shape}, we compare these quantities measured from
subhalos selected with $\vtoday$ and $\vpeak$, as well as from dark
matter particles.  We find that dark matter particles tend to have a
more prolate distribution than subhalos, and that subhalos show a
higher line-of-sight scatter of velocity dispersion.  Subhalos
selected with $\vpeak$ are slightly more elliptically distributed than
those selected with $\vtoday$.

\item {\em Formation history and subhalo properties.} We have
quantified the correlations between various subhalo properties and
halo formation history in Section~\ref{sec:cor} and in
Figure~\ref{fig:correlation_sub}.  The fraction of mass in subhalos
and the central dominance are both highly correlated with formation
time, late-time accretion rate, and concentration.  These correlations
have important implications for interpreting lensing-selected and
X-ray selected clusters.

\item {\em A fossil cluster.} Our sample includes a peculiar outlier,
Halo 572 (presented in Section~\ref{sec:572}), with exceptionally high
formation redshift, concentration, and central dominance.  It also has
exceptionally low late-time mass accretion rate, subhalo number, and
subhalo mass fraction.  This finding indicates that halos of distinct
formation history are likely to be distinguishable observationally, if
stringent selection criteria are used.

\item {\em Impact of tidal stripping on the occupation number of
subhalos.} In Section~\ref{sec:nocor}, we have demonstrated that the
subhalo number, when selected using $\vpeak$ (a more observationally
relevant property), does not correlate with formation time.  This is
in contrast to the result, shown previously and confirmed here, that
early-forming halos have fewer subhalos when selected with $\vtoday$.
We demonstrate that the correlation with the number of subhalos
selected with $\vtoday$ can be attributed to subhalo stripping and
insufficient resolution and is thus largely overestimated for cluster
satellite galaxies.  This finding implies that the assumption that
halo occupation number is independent of formation time at fixed mass
is likely to be a good one for luminosity or stellar mass selected
samples, and that the formation history of clusters is unlikely to be
directly inferred from the number of their satellite galaxies.

\end{enumerate}

In a forthcoming paper (H.-Y.~Wu et al., in preparation), we will
address the issue of the completeness of subhalo populations in detail
by comparing simulations of different resolutions directly with
observations.  We will also investigate the impact of completeness on
the measured velocity dispersion of subhalos.

The lack of a correlation between halo occupation number and formation
time provides support for halo occupation models that depend only on
halo mass, when the galaxy samples are selected by stellar mass or
luminosity.  However, our findings indicate that the halo occupation
is likely a function of selection; the halo occupation of
color-selected samples may depend on formation time.  This appears to
be consistent with some observational studies; e.g., the
luminosity-selected samples studied in \cite{TinkerConroy} did not
show evidence for trends with formation time, while evidence for this
dependence in samples selected by star formation rate was presented by
\cite{Tinker12}.

Although we have found that the formation history of clusters does not
manifest itself in the number of galaxies, correlation with formation
time still exists for subhalo mass fraction, central dominance, and
halo concentration.  These correlations are potentially observable in
targeted lensing programs like CLASH \citep{Postman12}, optical
follow-up programs for clusters detected by the South Pole Telescope
\citep{High12,Song12}, as well as the recent lensing mass calibration
for X-ray selected clusters by \cite{vdLinden12}.  These properties
together can indicate a system's state of relaxedness and can
potentially be combined to reduce the scatter in the observable--mass
relation for multi-wavelength surveys.

Finally, the dependence of subhalo statistics on the selection method
could potentially impact the prediction of galaxy clustering based on
the halo model.  For example, it is common to assume that the galaxy
number is described by a Poisson distribution, or that their spatial
distribution and velocities follow those of the dark matter particles
\citep[e.g.,][]{Zehavi11,Cacciato12}.  As we have shown, these
assumptions depend on the specific subhalo selection applied and on
the simulation resolution and are still uncertain.  Therefore, these
uncertainties will potentially limit the accuracy with which we can
predict the small-scale clustering and hence our ability to use it to
infer cosmological parameters \citep[see][]{WuHuterer13}.

\acknowledgements 

We thank Gus Evrard, Eduardo Rozo, Michael Busha, Matt Becker, and
Andrew Wetzel for many helpful suggestions and comments.  We also
thank the anonymous referee for many insightful comments.  We are
grateful to Michael Busha for providing the {\sc Carmen} simulation on
which the Rhapsody sample was based.  This work was supported by the
U.S. Department of Energy under contract numbers DE-AC02-76SF00515 and
DE-FG02-95ER40899 and by SLAC-LDRD-0030-12, and by Stanford University
through a Gabilan Stanford Graduate Fellowship to H.W. and a Terman
Fellowship to R.H.W.  O.H. acknowledges support from the Swiss
National Science Foundation (SNSF) through the Ambizione Fellowship.

\bibliographystyle{apj}
\bibliography{/Users/hao-yiwu/Dropbox/OrphanPaper/master_refs}
\end{document}